\documentclass[fleqn,usenatbib]{mnras}

\usepackage[T1]{fontenc}
\usepackage{ae,aecompl}

\usepackage{graphicx}	
\usepackage{amsmath}	
\usepackage{amssymb}	
\usepackage{float}
\usepackage{dblfloatfix}
\usepackage{subfigure}
\usepackage{blindtext}
\usepackage[dvipsnames]{xcolor}
\usepackage{enumitem}
\usepackage[colorinlistoftodos]{todonotes}
\usepackage[flushleft]{threeparttable}

\usepackage{indentfirst}
 
\usepackage{graphicx,amsmath}
\usepackage{color}
\usepackage{booktabs}
\usepackage{times}
\usepackage{float}
\usepackage{ulem}
\usepackage{multirow}
\usepackage{appendix}
\usepackage{longtable}
\usepackage{lscape}
\usepackage{supertabular}

\def\be{\begin{equation}}
\def\ee{\end{equation}}
\def\bea{\begin{eqnarray}}
\def\eea{\end{eqnarray}}

\title[abnormal quasars from \textit{Gaia}]
{Strange Quasar Candidates with Abnormal Astrometric Characteristics from \textit{Gaia} EDR3 and SDSS (SQUAB-II): Optical Identifications}

\author[Ji et al.]
{Xiang Ji$^{1}$\thanks{jixiang@shao.ac.cn},
Zhen-Ya Zheng$^{1,2}$\thanks{zhengzy@shao.ac.cn},
Qiqi Wu$^{1,2}$,
Ruqiu Lin$^{1,2}$,
P.T. Rahna$^{1}$,
Yingkang Zhang$^{1}$,
\newauthor
Shuairu Zhu$^{1,2}$,
Shilong Liao$^{1,2}$,
Zhaoxiang Qi$^{1,2}$,
Tao An$^{1,2,3}$
\\
$^{1}$Shanghai Astronomical Observatory, Chinese Academy of Sciences, 80 Nandan Road, Shanghai 200030, China\\
$^{2}$School of Astronomy and Space Sciences, University of Chinese Academy of Sciences, No. 19A Yuquan Road, Beijing 100049, People’s Republic of China \\
$^{3}$ Key Laboratory of Radio Astronomy and Technology, Chinese Academy of Sciences, A20 Datun Road, Chaoyang District, Beijing, 100101, P. R. China\\
}

\begin{document}
\label{firstpage}
\pagerange{\pageref{firstpage}--\pageref{lastpage}}
\maketitle

\begin{abstract} 
There are some strange quasars with multiple \textit{Gaia} detections or observed with abnormal astrometric characteristics, such as with large proper motions or significant astrometric noises. Those strange quasars could be potential candidates of quasar-star pairs, dual quasars (DQs), or lensed quasars (LQs). Searching for both DQs and LQs is of great importance in many fields of astrophysics. Here in this work, we select 143 SDSS spectroscopically confirmed quasars that have multiple \textit{Gaia} EDR3 detections within 1 arcsec of the SDSS quasar's position. We apply several optical identification methods to classify this sample. We firstly exclude 65 quasar-star pairs via their stellar features including their parallaxes and proper motions, stellar features in the SDSS spectra, or via the colour-colour diagram. Based on the spectral-fitting results, we find 2 DQ candidates, one of which presents a double-peaked [O III] emission line feature and the other shows a broad $H_{\beta}$ velocity offset ($\sim$ 870 km s$^{-1}$) relative to the [O III] $\lambda$5007 line. Via the colour difference method, we further find 56 LQ candidates with similar colours in their multiple images. We also cross-match 143 objects with the HST archive and find 19 targets with archival HST images. Our classification results of those 19 targets are mainly consistent with previous works.

\end{abstract}


\section{Introduction}
\label{sec:intro}

Quasars are regarded as point-like and distant sources with nearly zero proper motions, thus they could be used to establish a fundamental celestial reference frame and characterize the properties of \textit{Gaia} astrometric solution \citep[][]{Gaia2018b,Liao2019,LiaoPASP2021a,LiaoPASP2021b,GaiaCRF3}. However, there are some strange quasars with multiple \textit{Gaia} counterparts or observed with strange astrometric behaviours. For example, some strange quasars have been observed with large proper motions or significant astrometric noises, making them unsuitable for establishing celestial reference frames \citep[e.g.,][]{Wu2021,Wu2022,Souchay2022}. These strange quasars are interesting targets to reveal their structures and figure out the underlying  physical mechanisms. The anomaly could be explained by quasar-star pairs, dual quasars (DQs), or lensed quasars (LQs), of which the latter two are of great importance in many astrophysical fields.

Supermassive black hole binaries (SMBHBs) are expected to be abundant in the centers of many massive galaxies \citep[e.g.,][]{Volonteri2003,Chen2020} according to the hierarchical evolution models \citep[e.g.,][]{Begelman1980, Yu2002}. Searching for SMBHBs at different separations is important for our understanding of the merging process of galaxies, the formation and evolution of SMBHBs, and the origin of subsequent gravitational wave radiation \citep[e.g.,][]{Sesana2009, Chen2020}. Quasar pairs are progenitors of the SMBHB mergers. In particular, dedicated efforts have been put into searching for dual quasar candidates at a few kpc scales with double-peaked features of [O III] \citep[e.g.,][]{Liu2011,Fu2018}, although it has been proved only a small fraction of double-peaked targets are genuine dual quasars \citep[e.g.][]{Shen2011,Comerford2015}. However, there are a number of quasar pairs with separations from pc to kpc scales reported in the past few decades \citep[e.g.,][]{Liu2010b, Comerford2009, Shen2011, Ge2012, Koss2012,Comerford2015,Liu2018}. Among those pairs, there is only one SMBHB with a separation smaller than 10 pc discovered by the Very Long Baseline Array (VLBA) \citep{Rodriguez2006}, which indicates that Very Long Baseline Interferometry (VLBI) can be a powerful tool for confirming pc-scale SMBHBs \citep{An2018}. Along with the process of merger, quasar pairs would further become sub-parsec SMBHBs \citep[e.g.,][]{Yu2011}. However, sub-pc scale SMBHBs remain difficult to confirm. This is caused not only by the fact that these systems are too close to being spatially resolved \citep[e.g.,][]{Yu2002, Burke-Spolaor2011}, but also by the lack of distinct and unique features of these systems that can be used to confirm their duality \cite[e.g.,][]{Popovic2012, Graham2004}. Nevertheless, more than one hundred sub-pc SMBHB candidates have been found by using indirect signatures such as the periodical variation of the continuum or broad emission lines (BELs) \citep[e.g.,][]{Graham2015a, Graham2015b, Charisi2016, Li2016}, the optical/UV deficit in the continuum spectrum of a quasar \citep[e.g.,][]{Yan2015, Zheng2016}, or double-peaked/asymmetric profiles of BELs \citep[e.g.,][]{Tsalmantza2011, Eracleous2012, Shen2011, Liu2014, Li2016, Ji2021}. 

Gravitational lensed quasars are another type of typical strange quasars. Generally, they are useful tools to probe many important features in both quasars and lensing galaxies \citep[e.g.,][]{Ding2017,Stacey2018}, for example, revealing the geometry and kinematics of accretion disks \citep[e.g.,][]{Blackburne2011,Blackburne2015} or the broad line region of quasars \citep[e.g.,][]{Sluse2011}, inferring properties of the dark matter halos \citep[e.g.,][]{Oguri2004}, and even measuring the Hubble constant  \citep[e.g.,][]{Suyu2017}. 

Thanks to \textit{Gaia}, there are several works on searching for DQs and LQs from their  astrometric properties. With the release of \textit{Gaia} data release 2 (\textit{Gaia} DR2)\citep{Gaia2018a}, over one hundred newly detected LQs have been reported \citep[e.g.][]{Lemon2017,Lemon2018, Lemon2019,Lemon2021}. Besides,  the technique of machine learning has also been applied to discover the LQs \citep[e.g.][]{Krone-Martins2018,Ducourant2018,Delchambre2019} from \textit{Gaia} DR2. In addition, several new methods have also been proposed to search for DQs and LQs from \textit{Gaia}. For example, \cite{Shen2019} and \cite{Hwang2020} have proposed a new method of varstrometry, i.e., the astrometric jitter caused by flux variations from the multiple components in one system. Via this method, they have discovered several double quasars from the varstrometry selection \citep{Shen2021, Chen2022}. Another method is called the  \textit{Gaia} Multi Peak (GMP), i.e., the multiple peaks leading to another peak in the \textit{Gaia} point-spread function \citep[][]{Mannucci2022}. It has been proved with high efficiency in finding compact systems with separations between the multiple, point-like components in the range of $\sim$ 0.1\arcsec to 0.7\arcsec.  \citet[][]{Makarov2022ApJ} also provide a list of 44 candidate double or multiple AGNs with excess proper motions from \textit{Gaia} EDR3. As a consequence, the high-quality astrometric measurements from \textit{Gaia} provide us with a good opportunity to search for DQs and LQs systematically.

However, there is no systematic search for strange quasars from their astrometric behavior. The high-quality astrometric measurements from \textit{Gaia} have not been carefully considered for those strange quasars. In SQUAB-I \citep{Wu2022}, we have compiled a sample of strange quasars based on \textit{Gaia} EDR3 and SDSS DR16Q datasets. Those targets can be divided mainly into two categories. The first category consists of targets with more than one matched detection from \textit{Gaia} within a 1\arcsec radius of the SDSS's position. The second category consists of SDSS quasars with abnormal astrometric properties measured from \textit{Gaia}, for example, with significant astrometric noises. The present work (SQUAB-II) further examines their structures and makes classifications of those strange quasars in the first category. We have developed several methods to get rid of contamination of the quasar-star pairs and select DQ candidates and LQ candidates for further spectroscopic confirmation. 

The paper is organized as follows. In Section ~\ref{sec:data}, we briefly introduce the data used and the target selection strategy. Methods to select the potential quasar-star pairs, LQ candidates, and DQ candidates are described in Section ~\ref{sec:method}. In Section ~\ref{sec:results}, we present optical confirmation of our selected LQ or DQ candidates with the help of the Pan-STARRS images. In Section~\ref{sec:discussion}, we first give brief discussions about our target selection methods, describe the possibility of quasar-star superpositions, and judge the extra photometric effects from nearby objects. We further check the HST archival for available high-resolution HST images, and also make comparisons with previous works. Finally, the main conclusions from this work are summarized in Section ~\ref{sec:conclusion}. 

\begin{figure}
\centering
\includegraphics[width=0.5\textwidth]{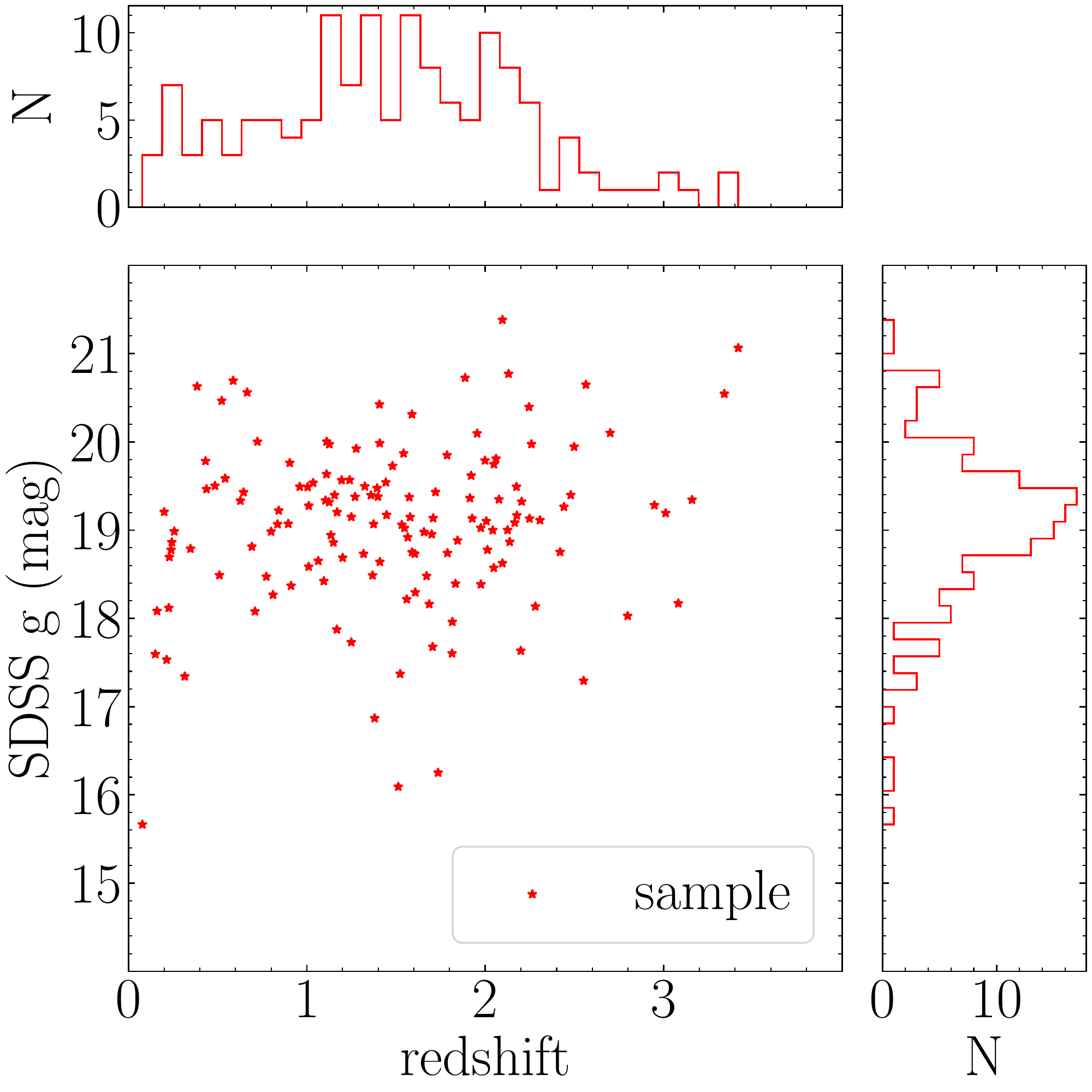}
\caption{Distributions of the redshift and the SDSS g magnitude for all the targets in our sample. The top and the right histograms show the number density of the distributions for the redshift and the SDSS g magnitude, respectively.
}
\label{fig:f1}
\end{figure} 

\section{Data and sample}
\label{sec:data}

\textit{Gaia} is an all-sky space astrometric mission \citep{Gaia2016} with the most precise detection ability currently. It provides high-quality astrometric measurements including the proper motion and parallax measurements, and the colour information of three passbands, i.e. \textit{G}, \textit{$\rm G_{BP}$}, and \textit{$\rm G_{RP}$}. For a target at {\it G}=20 mag, the photometric uncertainties for \textit{G}, $\rm G_{BP}$ and \textit{$\rm G_{RP}$} are about 6, 108, and 52 mmag, respectively. The {\it G} band magnitude limit is about 20.7 mag and the highest angular resolution is $\sim$ 0\farcs18. As a consequence, the high-quality astrometric measurements from \textit{Gaia} allow us to exclude stellar contaminates and identify candidates of both LQs and DQs. 

\begin{landscape}
\begin{table}
    \caption{Basic properties of the known LQs and DQs (candidates). The columns $\rm N_{image}$ and $\rm N_{Gaia}$ represent the number of the cores in one system and that detected by \textit{Gaia}, respectively.}
    \label{table:tb1}
    \begin{tabular*}{1.4\textwidth}{@{\extracolsep{\fill}}cccccccccccc}
    \hline
    \textbf{name} &  \textbf{redshift} & \textbf{SDSS $g$} & \textbf{ra$_\textrm{Gaia}$} & \textbf{dec$_\textrm{Gaia}$} 
    & \textbf{$G$} & \textbf{$G_\textrm{BP}$} &\textbf{$ G_\textrm{RP}$}  & \textbf{$\rm N_{image}$} & \textbf{$\rm N_{Gaia}$} & \textbf{class} & \textbf{reference} \\ 
    \hline \hline 
081331.28+254503.0 & 1.512    & 16.092 $\pm$ 0.003   & 123.38030149239 & 25.75086321438 & 16.338 $\pm$ 0.012    & 16.211 $\pm$ 0.010      & 15.174 $\pm$ 0.008     & 4 & 2     & LQs           & [1][2]   \\
                   &        &         & 123.38054144472 & 25.75079410426 & 17.940 $\pm$ 0.007    &      &          &       &               &    &   \\
111816.94+074558.2 & 1.736    & 16.252 $\pm$ 0.004   & 169.57066749217 & 7.7662504694   & 17.086 $\pm$ 0.011    & 16.481 $\pm$ 0.031     & 15.853 $\pm$ 0.021     & 4 & 4     & LQs           & [3][4]  \\
                   &              &         & 169.57062595971 & 7.76612498888  & 17.211 $\pm$ 0.009    & 16.611 $\pm$ 0.043     & 15.966 $\pm$ 0.033     &   &       &               &       \\
                   &              &         & 169.57025416162 & 7.7666884210447  & 18.528 $\pm$ 0.006    & 18.574 $\pm$ 0.074     & 17.979 $\pm$ 0.096     &   &       &               &       \\
                   &              &         & 169.570157868845 &
7.7661434790373  & 18.866 $\pm$ 0.008    &      &     &   &       &               &       \\
091301.02+525928.8 & 1.380     & 16.868 $\pm$ 0.004   & 138.25417691679 & 52.99138630842 & 16.484 $\pm$ 0.005    & 16.470  $\pm$ 0.016     & 15.552 $\pm$ 0.018     & 2 & 2     & LQs           & [5]     \\
                   &               &         & 138.25463214963 & 52.99124835768 & 17.068 $\pm$ 0.005    & 16.687 $\pm$ 0.270      & 15.753 $\pm$ 0.294     &   &       &               &       \\
141546.24+112943.4 & 2.551    & 17.293 $\pm$ 0.004   & 213.94261159532 & 11.49531201636 & 17.569 $\pm$ 0.015    & 16.788 $\pm$ 0.037     & 16.022 $\pm$ 0.025     & 4 & 3     & LQs           & [6]     \\
                   &              &         & 213.94281909923 & 11.49536505201 & 17.574 $\pm$ 0.016    & 17.002 $\pm$ 0.109     & 16.057 $\pm$ 0.114     &   &       &               &       \\
                   &              &         & 213.9425258972 &
11.4955317232 & 17.638 $\pm$ 0.019    &     &      &   &       &               &       \\  
163348.98+313411.9 & 1.523    & 17.371 $\pm$ 0.005   & 248.4540730856  & 31.5700034202  & 17.432 $\pm$ 0.004    & 17.432 $\pm$ 0.011     & 16.795 $\pm$ 0.010      & 2 & 2     & LQs           & [7]     \\
                   &              &         & 248.45424348917 & 31.56989117914 & 19.132 $\pm$ 0.006    &             &              &   &       &               &       \\
095122.57+263513.9 & 1.249    & 17.729 $\pm$ 0.005   & 147.84403947441 & 26.58723286072 & 17.490  $\pm$ 0.006    & 17.522 $\pm$ 0.019     & 16.855 $\pm$ 0.015     & 2 & 2     & LQs           & [8]     \\
                   &                 &         & 147.84431797953 & 26.5870561662  & 19.199 $\pm$ 0.009    &         &         &   &       &               &  \\
091127.61+055054.1 & 2.798    & 18.028 $\pm$ 0.006   & 137.86513574363 & 5.84841020213  & 18.870  $\pm$ 0.014    & 18.397 $\pm$ 0.028     & 17.762 $\pm$ 0.016     & 4 & 4     & LQs           & [9]     \\
                   &              &         & 137.86423192750 &
5.848517226935 & 19.875 $\pm$ 0.008     &   20.073 $\pm 0.092$         &    19.423  $\pm 0.111$    &   &       &      &    \\
                   &              &         & 137.86505887342 & 5.84829857367  & 19.508 $\pm$ 0.020     &             &         &   &       &      &    \\
                   &              &         & 137.86505738812 &
5.848563639185  & 19.745 $\pm$ 0.016     &             &         &   &       &      &    \\
024634.09-082536.1 & 1.686    & 18.162 $\pm$ 0.007   & 41.64211559961  & -8.42672188068 & 16.929 $\pm$ 0.014    & 16.959 $\pm$0.044     & 16.502 $\pm$ 0.032     & 2 & 2     & LQs           & [10]    \\
                   &               &         & 41.64186218579  & -8.42654907382 & 18.861 $\pm$ 0.011    &         &         &   &       &         &  \\
133222.62+034739.9 & 1.447    & 19.171 $\pm$ 0.008   & 203.0945132929  & 3.79432898831  & 19.934 $\pm$ 0.010     & 19.437 $\pm$ 0.130      & 18.205 $\pm$ 0.076     & 2 & 2     & LQs           & [11]    \\
                   &                  &         & 203.09423472652 & 3.79446633063  & 20.295 $\pm$ 0.023    & 19.331 $\pm$ 0.044     & 18.015 $\pm$ 0.023     &   &       &               &       \\
074653.04+440351.3 & 2.006    & 19.102 $\pm$ 0.008   & 116.72118927293 & 44.06422019002 & 19.472$\pm$  0.009    & 19.019 $\pm$ 0.065     & 18.287$\pm$ 0.032     & 2 & 2     & LQs           & [12]    \\
                   &              &         & 116.72082301636 & 44.06436462402 & 19.624 $\pm$ 0.010     & 19.203 $\pm$ 0.098     & 18.369 $\pm$ 0.043     &   &       &               &       \\
102111.01+491330.3 & 1.721    & 19.431 $\pm$ 0.014   & 155.29599915159 & 49.22503654914 & 19.525 $\pm$ 0.009    & 19.423 $\pm$0.036     & 18.522 $\pm$ 0.026     & 2 & 2     & LQs           & [13]    \\
                   &                &         & 155.2956378427  & 49.22525313164 & 20.998 $\pm$ 0.025    &         &         &   &       &               &       \\
112818.49+240217.4 & 1.608    & 18.295 $\pm$ 0.007   & 172.07698911386 & 24.03815267311 & 18.531$\pm$ 0.008    & 18.383 $\pm$ 0.034     & 17.598 $\pm$ 0.031     & 2 & 2     & LQs           & [14][15] \\
                   &              &         & 172.07717757309 & 24.03828375358 & 19.317 $\pm$ 0.008    &         &        &   &       &      &  \\
234330.58+043557.9 & 1.605    & 18.732 $\pm$ 0.006   & 355.87759501346 & 4.5993938917   & 18.635 $\pm$ 0.011    & 18.654 $\pm$ 0.046 & 17.489 $\pm$ 0.034  & 2 & 2  & LQs  & [18]  \\
                   &          &         & 355.87727347014 & 4.59951311798  & 18.961 $\pm$ 0.008    &     &       &   &      &      &       \\
133401.39+331534.3 & 2.419    & 18.752 $\pm$ 0.007   & 203.50578096139 & 33.25947865395 & 19.597$\pm$ 0.006    & 19.115 $\pm$ 0.03      & 18.622 $\pm$ 0.046     & 2 & 2     & LQs           & [16]    \\
                   &              &         & 203.50594425795 & 33.25966704236 & 19.793 $\pm$  0.007    & 19.047 $\pm$  0.044     & 18.628 $\pm$ 0.053     &   &       &               &       \\
132128.67+541855.5 & 2.259    & 19.974 $\pm$ 0.019   & 200.36940900851 & 54.31544057296 & 20.265 $\pm$0.013    & 20.185 $\pm$ 0.097     & 19.477 $\pm$ 0.070      & 2 & 2     & LQ candidate & [16]    \\
                   &              &         & 200.36978382582 & 54.31540192645 & 20.943 $\pm$ 0.033    &         &       &   &       &       &       \\
162501.98+430931.6 & 1.657    & 18.978 $\pm$ 0.009   & 246.25829345634 & 43.15872055958 & 19.233 $\pm$ 0.011    & 19.007 $\pm$ 0.034     & 18.369 $\pm$ 0.042     & 2 & 2     & LQ candidate & [16]    \\
                   &              &         & 246.25818775771 & 43.15885994875 & 20.016 $\pm$ 0.010     &        &        &   &       &             &     \\
105926.43+062227.4 & 2.199    & 17.633 $\pm$0.005   & 164.86015095584 & 6.37420498672  & 17.462 $\pm$ 0.015    & 17.205 $\pm$ 0.024     & 16.538 $\pm$0.017     & 2 & 2     & LQ candidate & [16][17] \\
                   &              &         & 164.86009996452 & 6.37436131019  & 17.946 $\pm$ 0.020     &      &         &   &       &          &      \\
212243.01-002653.6 & 1.975    & 19.025 $\pm$  0.010    & 320.67921837391 & -0.44828161549 & 19.270  $\pm$  0.006    & 19.078 $\pm$ 0.043     & 18.243 $\pm$ 0.038     & 2 & 2     & LQ candidate & [19]    \\
                   &           &         & 320.67912428067 & -0.44816230327 & 19.860  $\pm$ 0.009    &         &     &   &       &             &       \\
205752.47+000635.2 & 1.135    & 18.943 $\pm$ 0.009   & 314.46863940567 & 0.10978924318  & 18.718 $\pm$ 0.004    & 18.744 $\pm$ 0.016     & 18.201 $\pm$0.031     & 2 & 2     & LQ candidate & [19]    \\
                   &                  &         & 314.46874170038 & 0.11004439111  & 20.407 $\pm$ 0.007    &      &         &   &       &              &       \\
084129.77+482548.3 & 2.948    & 19.283 $\pm$ 0.010    & 130.37404617611 & 48.43013479612 & 19.478 $\pm$ 0.011    & 19.19  $\pm$ 0.038     & 18.454 $\pm$ 0.038     & 2 & 2     & DQs           & [20][21] \\
                   &              &         & 130.37417633919 & 48.43005444844 & 19.78 $\pm$ 0.037   & 19.345$\pm$ 0.091     & 18.267$\pm$ 0.115     &   &       &               &       \\
084710.40-001302.6 & 0.628    & 19.333 $\pm$ 0.010    & 131.79334744852 & -0.21735281015 & 19.713 $\pm$ 0.009    & 19.245 $\pm$ 0.037     & 18.157 $\pm$ 0.027     & 2 & 2     & DQs           & [22][23] \\
                   &              &         & 131.79349676225 & -0.21757556956 & 20.304 $\pm$ 0.012    &        &       &   &       &               &       \\
123915.40+531414.6 & 0.200      & 19.206 $\pm$ 0.007   & 189.8143793831  & 53.2374973296  & 20.399 $\pm$ 0.023    & 18.719 $\pm$ 0.045     & 17.053 $\pm$ 0.019     & 2 & 2     & DQ candidate & [24]    \\
                   &               &         & 189.81404358851 & 53.23722292116 & 21.414 $\pm$ 0.041    & 18.615 $\pm$0.090      & 17.018 $\pm$0.048     &   &       &               &     \\
    \hline\hline
    \end{tabular*}
    \scriptsize
    \begin{tablenotes}
    \item References:
    [1] \citet{Reimer2002};   [2]\citet{Jackson2015};
    [3] \citet{Young1981};    [4] \citet{Chiba2005};
    [5] \citet{Lehar2006};    [6] \citet{Magain1988};
    [7] \citet{Morgan2001};   [8] \citet{Schechter1998};
    [9] \citet{Burud1998};    [10] \citet{Inada2005};
    [11] \citet{Morokuma2007}; [12] \citet{Inada2007};
    [13] \citet{Pindor2006};  [14] \citet{Inada2014}; 
    [15] \citet{Agnello2018};  [16] \citet{Lemon2017};
    [17] \citet{Ducourant2018}; [18] \citet{Krone-Martins2019}
    [19] \citet{Lacki2009};     [20]\citet{Shen2021}; 
    [21] \citet{Mannucci2022};  [22]\citet{Inada2008}; 
    [23] \citet{Silverman2020};  [24]\citet{Smith2010};
    \end{tablenotes}
\end{table}
\end{landscape}

Our target sample consists of sources from \textit{Gaia} EDR3 \citep[][]{Gaia2021} and our parent quasar sample are spectroscopic identified quasars from SDSS DR16 \citep[][]{Jonsson2020,Lyke2020}. In this work, we focus on the sources identified as one quasar in SDSS but with two or more \textit{Gaia} detections within 1\arcsec\ \footnote{Given the fiber diameter of 3\arcsec\ for the SDSS spectra, we require separation radius of 1\arcsec\ to search for more close pairs.} of the SDSS quasar's position. Our sample consists of 143\footnote{In our paper I, there are 155 targets fulfilling the criteria. However, we find 12 of them labeled as 'stars' according to the SDSS spectra, although they are still in the SDSS DR16q catalog. We get rid of those targets in this work.} SDSS quasars in total. The sample selection method has been described in SQUAB-I in detail. 

The distributions of the SDSS g magnitudes and the spectroscopic redshifts from SDSS for all these strange SDSS quasars are shown in Figure \ref{fig:f1}. Furthermore, among the 143 SDSS quasars, there are 22 SDSS quasars reported by previous works, of which 14 are known LQs, 5 are LQ candidates, 2 are known DQs, and 1 is a DQ candidate (see Table \ref{table:tb1} for details). Those known LQs/DQs would serve as a reference when carrying out the colour difference analysis in the following section.

\section{Method}
\label{sec:method}

In this section, we mainly describe our procedures to classify the strange SDSS quasars in detail. Generally, our procedures can be divided into three steps. Firstly, we get rid of  quasar-star pairs with distinct stellar features in Section~\ref{sec:method:stellar}. Then in Section~\ref{sec:method:emission}, we further deal with the SDSS spectra and search for double-peaked or offset emission line profiles, which could be indirect features of the existence of DQs. Finally, we search for LQ candidates with similar colour differences between their multiple detections from \textit{Gaia} in Section~\ref{sec:method:colordiff}. 

\subsection{Stellar Features}
\label{sec:method:stellar}
We regard targets as quasar-star pairs once it fulfills one of the following three conditions. Firstly, we select targets of which one or more detections from \textit{Gaia} in one system have proper motions or parallaxes greater than 5$\sigma$ significance, i.e.,

\begin{equation}
|\varpi /\sigma_{\varpi}|>5 \ or\  |\mu/\sigma_{\mu}| >5 , 
\label{equ:proper motion}
\end{equation}
where $\varpi$, $\sigma_{\varpi}$, $\mu$ and $\sigma_{\mu}$ represent the parallax, parallax error, the total proper motion, and the total proper motion error, respectively. Secondly, their SDSS spectra could be well explained by quasar-star superpositions as suggested by \cite{shen2022}, using a method called the principal component analysis (PCA). Thirdly, we use the colour-colour diagram as proposed in \citep{Bailer-Jones2019,Gaia2022} to select objects with similar colors as stars. Considering that only a small portion of those targets have measurements for both detections of \textit{$\rm G_{BP}$} and \textit{$\rm G_{RP}$} passbands from \textit{Gaia} in one system, we divide them into two sub-samples. Those targets with both detections of \textit{$\rm G_{BP}$} and \textit{$\rm G_{RP}$} passbands are labeled as sub-sample A. Those with only one detection are labeled as sub-sample B. We further pick out targets of which one or more \textit{Gaia} detections have a larger confidence level to be a star than to be a quasar or galaxy in the colour-colour diagram. See Section ~\ref {sec:stars} for more details. With the above three criteria, we have identified 65 quasar-star pairs, which are excluded in the following analysis.

\subsection{Emission line features} 
\label{sec:method:emission}

Generally, the double-peaked narrow emission line of [O III] could be an indirect signature of the presence of DQs and a series of efforts have been put into searching for DQs through this way \citep[e.g.][]{Comerford2012,Liu2010a,Smith2010}. On the other aspect, the offset line properties could also be a possible signature of galaxy merger and hence imply the existence of offset active galactic nuclei (AGNs) or DQs \citep[e.g.][]{Comerford2013}. Usually, the offset could be manifested as the difference between the broad/narrow emission lines and the quasar rest frame (the host galaxy) \citep[e.g.][]{Eracleous2012,Comerford2014,Tsalmantza2011} or the broad emission line in respect to the narrow emission line \citep[e.g.][]{shen2013}. 

Considering the different redshifts for the targets in our sample, we search for the potential double-peaked features of broad $\rm H_{\beta}$, $\rm  H_{\alpha}$, $\rm Mg \ II$, and $ \rm C \ IV$ emission lines as well as the narrow [O III] emission line. We first visually inspect the spectra and try to find the potential double-peaked features. In particular, there are 17 targets in total in our sample with broad $\rm H_{\beta}$ emission lines and [O III] narrow emission lines due to redshift restriction. Then we use one or two Gaussians to fit both the broad and narrow emission lines \citep{Ge2012,Feng2021}. Then the line offset between the broad $\rm H_{\beta}$ line relative to the [O III] $\lambda$5007 line can be figured out based on the spectra fitting results. 

\subsection{Colour difference}
\label{sec:method:colordiff}
Generally, the colour difference between lensed images in the same LQ system is usually very small. For the remaining targets in sub-sample A, we derive the colour difference of $\rm (G_{BP}-G_{RP})_{d1} - (G_{BP}-G_{RP})_{d2}$ for them, where $\rm (G_{BP}-G_{RP})_{d1}$ and $\rm (G_{BP}-G_{RP})_{d2}$ represent the colour for the brighter detection and the dimmer detection in the same system, respectively. Due to the lack of the \textit{$\rm G_{BP}$} and \textit{$\rm G_{RP}$} measurements for the dimmer detection for the remaining targets in sub-sample B, we thus combine the $\rm G_{BP}$, and $\rm G_{RP}$ of the brighter detection together with the SDSS {\it g} and {\it i} magnitudes to obtain the colour difference, i.e., \textbf{$\rm (G_{BP}-G_{RP})_{d1} - (g-i)_{tot}$}. The colour information \textbf{$\rm (g-i)_{tot}$} from SDSS could represent the color of the system as a whole. 

\begin{figure}
\centering
\includegraphics[width=0.5 \textwidth]{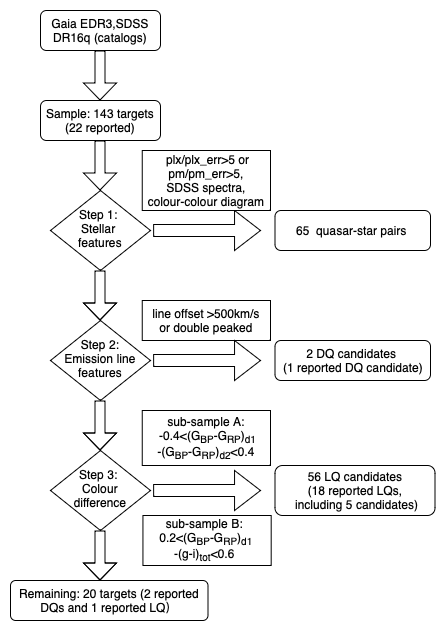}
\caption{Flowchart of the classification procedures of the targets in our sample. Firstly we compile a sample of SDSS quasars with multiple detections from \textit{Gaia} EDR3. The details of each classification criteria are described in Section ~\ref{sec:method}. We also mark the numbers of the selected targets in each step.
}
\label{fig:f2}
\end{figure}

The above classification procedures are shown in Figure ~\ref{fig:f2}. Based on these steps, we select 65 quasar-star pairs, 2 DQ candidates (1 previously reported), and 56 LQ candidates (18 previously known LQs including 5 candidates). However, there are 20 targets that can not be specified or classified based on the available data via our methods. We present the detailed analysis and results in section ~\ref{sec:results}.

\begin{figure}
\centering
\includegraphics[width=0.5\textwidth]{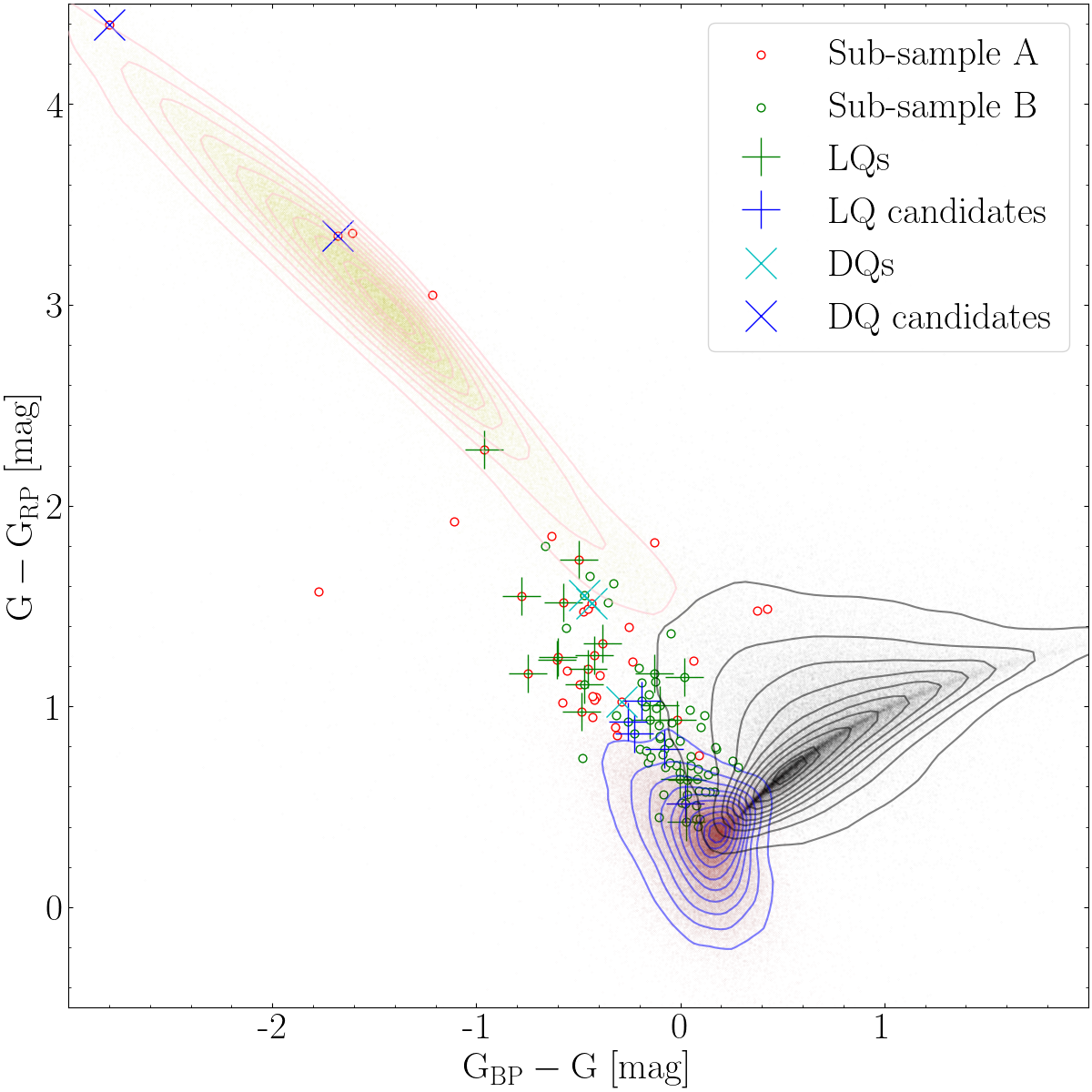}
\caption{Colour-colour diagram for all the \textit{Gaia} detections with measurements of magnitudes of \textit{G}, $\rm G_{BP}$, and $\rm G_{RP}$ passbands in our sample. The red and green circles represent for the \textit{Gaia} detections from sub-sample A and sub-sample B, respectively. For comparison, we have shown the colour-colour diagram from \citet{Gaia2022} as the translucent background in the above figure. The blue, black, and yellow parts in the background represent the quasars, stars, and galaxies, respectively. We also mark the \textit{Gaia} detections from the known LQs, LQ candidates, DQs, and DQ candidates in our sample with green pluses, blue pluses, cyan crosses, and blue crosses, respectively.
}
\label{fig:f3}
\end{figure}

\section{Results}
\label{sec:results}

In this section, we present the main classification results of our sample. Firstly, we get rid of the quasar-star pairs based on the method described in Section \ref {sec:method}. Secondly, we show the spectra fitting results of the discovered 2 DQ candidates. In the following, we single out 56 LQ candidates and show the optical confirmation with the help of the Pan-STARRS images of those candidates. Finally, we show the distributions of the separations as a function of the redshift for the quasars in our sample.

\newpage
\subsection{Quasar-star pairs}
\label{sec:stars}
Based on the high-quality astrometric measurements from \textit{Gaia}, we first get rid of targets in our sample of which one or more detections from \textit{Gaia} have large proper motion/parallax measurements. We find 55 systems in total. Those systems are more likely to be quasar-star pairs. In a recent paper \citep{shen2022}, they used a PCA method to get rid of quasar-star superpositions. After cross-matched with our 143 targets, we find 8 targets in common, which have already been selected by the proper motion/parallax criteria.

According to the colour-colour diagram as proposed in \cite{Bailer-Jones2019} and \cite{Gaia2022}, stars, quasars and galaxies are located in the different regions in the $\rm (G-G_{RP})$ versus $\rm (G_{BP}-G)$ colour-colour diagram, which could help us to distinguish different types of the targets in our sample. As shown in Figure \ref{fig:f3}, most of our targets locate in the region of quasars and extend to the region of galaxies, which takes on similar features of the distributions for the LQs identified in the literature. Still, there are a small fraction of detections from \textit{Gaia} with higher probabilities of being stars than that being quasars or galaxies. We further select 10 systems according to the colour-colour diagram. Based on all the above analyses, we identified 65 quasar-star pairs. We exclude those systems in the following analysis.

\begin{figure*}
\centering
\includegraphics[width=1.0 \textwidth]{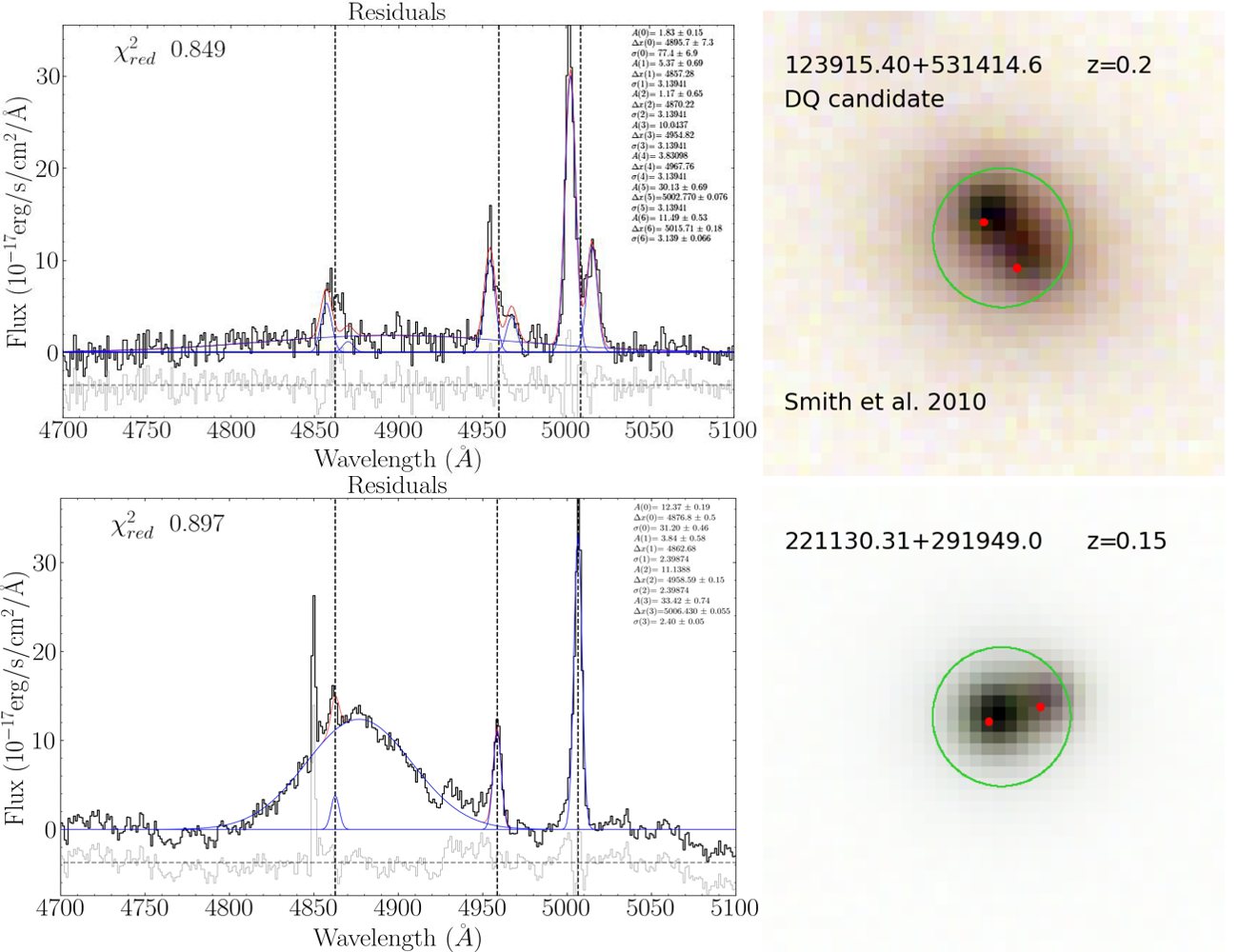}
\caption{Spectra fitting results and the optical identifications of the two DQ candidates we selected from our sample. The top and bottom panels are for the target SDSS J123915.40+531414.6 and SDSS J221130.31+291949.0, respectively. Left panels: the black, red, and blue lines are for the observational data, model, and Gaussians of the broad/narrow components. The values of the reduced $\chi$-square and the best-fitting parameters are marked in the upper left and right, respectively. Right panels: Pan-STARRS composite colour images (Pan-STARRS i band in red, Pan-STARRS r band in green, and Pan-STARRS g band in blue. We invert the colormap for the sake of presentation.) for the selected DQ candidates. For each image, the size is 10\arcsec\  $\times$ 10\arcsec\ , with the source id and the redshift marked in the upper left and upper right corners, respectively. The center of the green circle in each panel is from the SDSS coordinates, with a radius of 1\farcs5. In addition, we also mark the detections from \textit{Gaia} in red dots in each panel. 
}
\label{fig:f4}
\end{figure*}

\begin{figure*}
\centering
\includegraphics[width=1.0\textwidth]{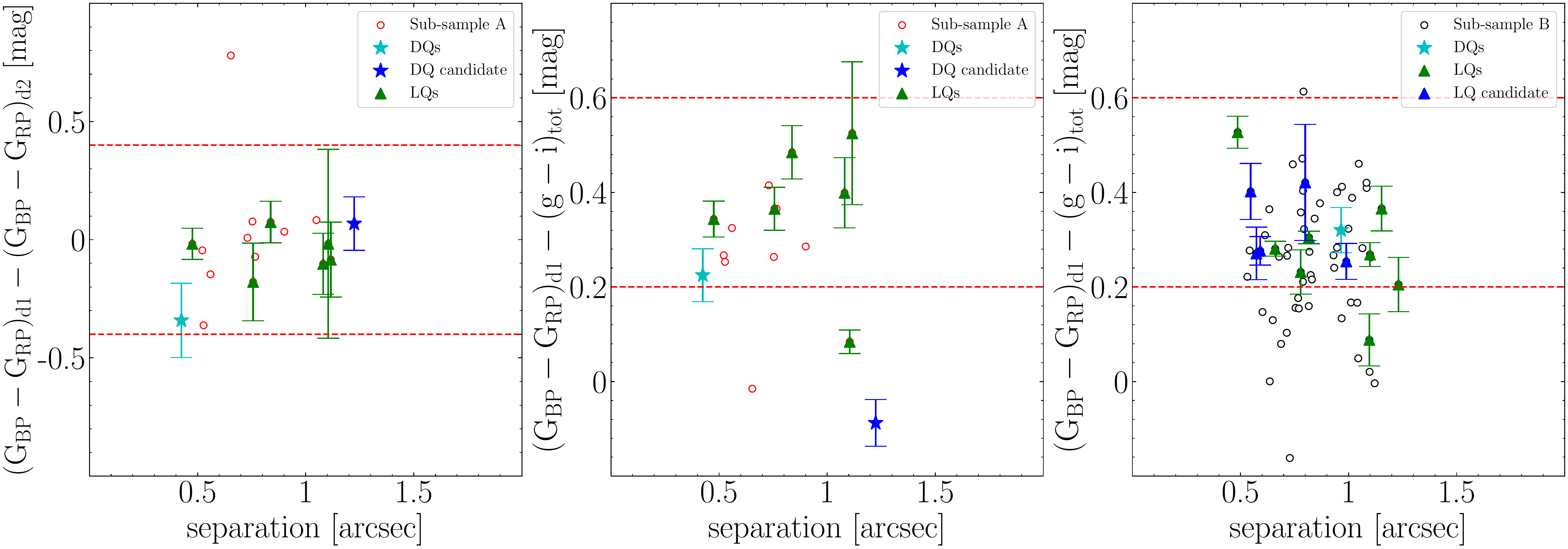}
\caption{Left panel: colour differences of the $\rm (G_{BP}-G_{RP})_{d1} - (G_{BP}-G_{RP})_{d2}$ as a relationship of the separations for the remaining targets in sub-sample A. The red empty circles, cyan stars, blue stars and green triangles are for the targets, the known DQs, the DQ candidates, and the known LQs, respectively. We also display the error bars for the reported DQs, DQ candidates, and LQs. The red dotted lines are the reference lines that indicate the range we select for the LQ candidates. Middle panel: colour differences of $\rm (G_{BP}-G_{RP})_{d1} -(g -i)_{tot}$ as a relationship of the separations for the remaining targets in sub-sample A. Legend are similar to that in the left panel. Right panel: similar to that in the middle panel but for the remaining targets in sub-sample B. The black empty circles, cyan stars, green triangles, and blue triangles are for the targets, the known DQs, known LQs and LQ candidates in sub-sample B, respectively. The red dotted lines are the reference lines that indicate the range we select for the LQ candidates in this sub-sample. 
}
\label{fig:f5}
\end{figure*}

\subsection{DQ candidates}

We select DQ candidates mainly via their spectral properties. There is one target, SDSS J123915.40+531414.6, with a distinct double-peaked emission line feature which has already been reported by \citet{Smith2010}. We use one Gaussian and two Gaussians to fit the broad and narrow components, respectively. The best fitting results are shown in the top left panel of Figure \ref{fig:f4}. We further make optical confirmations via the Pan-STARRS composite colour image with the \textit{Gaia} detections superimposed, which has not been considered before. However, due to the limited resolution of the Pan-STARRS image, the structure of this target is still elusive. 

We also find one target, SDSS J221130.31+291949.0, with a distinct line offset of the broad $H_{\beta}$ emission line with respect to the narrow [O III] $\lambda$5007 emission line. We use one Gaussian to fit both the broad and narrow emission lines and obtain the positions of the peaks for the emission lines. The best-fitting results and the optical confirmation are shown in the bottom left and right panels of Figure \ref{fig:f4}, respectively. Via the fitting results, we estimate the velocity offset to be about $\sim$ 870 km s$^{-1}$. Considering the line offset property as well as the two detections from \textit{Gaia}, this target is a potential DQ candidate. Although, we can't exclude the possibility of alternative explanations, like a recoiling black hole \citep[]{Eracleous2012}. We are carrying up follow-up observations to reveal its nature.

\begin{table*}
\caption{Basic properties of the LQ candidates from sub-sample A. The known LQs have been labeled with stars in the upper right of their source ids.}
\centering
\begin{tabular}{lllllllll} 
\hline\hline
\textbf{name} &  \textbf{redshift} & \textbf{SDSS $g$} & \textbf{ra$_\textrm{Gaia}$} & \textbf{dec$_\textrm{Gaia}$} & \textbf{$G$} & \textbf{$\rm G_\textrm{BP}$} &\textbf{$\rm G_\textrm{RP}$} \\  \hline

111816.94+074558.2$^{*}$ & 1.736   & 16.252 $\pm$ 0.004   & 169.57066749217 & 7.7662504694   & 17.086 $\pm$ 0.011    & 16.481 $\pm$ 0.031     & 15.853 $\pm$ 0.021 \\
                   &              &         & 169.57062595971 & 7.76612498888  & 17.211 $\pm$ 0.009    & 16.611 $\pm$ 0.043     & 15.966 $\pm$ 0.033  \\
                   &              &         & 169.57025416162 & 7.7666884210447  & 18.528 $\pm$ 0.006    & 18.574 $\pm$ 0.074     & 17.979 $\pm$ 0.096 \\
                   &              &         & 169.570157868845 & 7.7661434790373  & 18.866 $\pm$ 0.008     &     &  \\
091301.02+525928.8$^{*}$ & 1.380     & 16.868 $\pm$ 0.004   & 138.25417691679 & 52.99138630842 & 16.484 $\pm$ 0.005    & 16.470  $\pm$ 0.016     & 15.552 $\pm$ 0.018     \\
                   &               &         & 138.25463214963 & 52.99124835768 & 17.068 $\pm$ 0.005    & 16.687 $\pm$ 0.270      & 15.753 $\pm$ 0.294   \\
141546.24+112943.4$^{*}$ & 2.551    & 17.293 $\pm$ 0.004   & 213.94261159532 & 11.49531201636 & 17.569 $\pm$ 0.015    & 16.788 $\pm$ 0.037     & 16.022 $\pm$ 0.025   \\
                   &              &         & 213.94281909923 & 11.49536505201 & 17.574 $\pm$ 0.016    & 17.002 $\pm$ 0.109     & 16.057 $\pm$ 0.114    \\
                   &              &         & 213.9425258972 &
11.4955317232 & 17.638 $\pm$ 0.019    &     &       \\  
133222.62+034739.9$^{*}$ & 1.447    & 19.171 $\pm$ 0.008   & 203.0945132929  & 3.79432898831  & 19.934 $\pm$ 0.010     & 19.437 $\pm$ 0.130      & 18.205 $\pm$ 0.076    \\
                   &                  &         & 203.09423472652 & 3.79446633063  & 20.295 $\pm$ 0.023    & 19.331 $\pm$ 0.044     & 18.015 $\pm$ 0.023   \\
074653.04+440351.3$^{*}$ & 2.006    & 19.102 $\pm$ 0.008   & 116.72118927293 & 44.06422019002 & 19.472 $\pm$  0.009    & 19.019 $\pm$ 0.065     & 18.287 $\pm$ 0.032  \\
                   &              &         & 116.72082301636 & 44.06436462402 & 19.624 $\pm$ 0.010     & 19.203 $\pm$ 0.098     & 18.369 $\pm$ 0.043  \\
133401.39+331534.3$^{*}$ & 2.419    & 18.752 $\pm$ 0.007   & 203.50578096139 & 33.25947865395 & 19.597 $\pm$ 0.006    & 19.115 $\pm$ 0.030      & 18.622 $\pm$ 0.046    \\
                   &              &         & 203.50594425795 & 33.25966704236 & 19.793 $\pm$  0.007    & 19.047 $\pm$  0.044     & 18.628 $\pm$ 0.053     \\
081246.41+200729.9 & 1.480     & 19.638 $\pm$ 0.013   & 123.19338721538 & 20.12504974558 & 20.166 $\pm$ 0.014    & 19.753 $\pm$ 0.080      & 19.119 $\pm$ 0.070      \\
                   &      &   & 123.19339549084 & 20.12490340453 & 20.436 $\pm$ 0.018    & 19.962 $\pm$ 0.093     & 18.965 $\pm$ 0.237     \\
080009.98+165509.4 & 0.709    & 17.883 $\pm$ 0.006   & 120.04153952266 & 16.91934178839 & 18.304 $\pm$ 0.015    & 17.996 $\pm$ 0.033     & 17.448 $\pm$ 0.024     \\
                   &     &   & 120.04170900144 & 16.91920386557 & 18.709 $\pm$ 0.005    & 18.153 $\pm$ 0.004     & 17.533 $\pm$ 0.067     \\
074800.55+314647.7 & 1.408    & 19.803 $\pm$ 0.015   & 117.00230747636 & 31.77985563261 & 20.033 $\pm$ 0.007    & 19.797 $\pm$ 0.062     & 18.811 $\pm$ 0.051     \\
                   &     &   & 117.00227692679 & 31.77999820592 & 20.319 $\pm$ 0.014    & 19.864 $\pm$ 0.190      & 18.833 $\pm$ 0.041     \\
092011.82+373223.7 & 0.385    & 19.680  $\pm$ 0.026   & 140.049490253   & 37.53993026155 & 21.228 $\pm$ 0.023    & 20.012 $\pm$ 0.161     & 18.177 $\pm$ 0.040      \\
                   &     &   & 140.04912272088 & 37.53992137658 & 21.530  $\pm$ 0.031    & 19.923 $\pm$ 0.023     & 18.170  $\pm$ 0.117     \\
091150.63+240545.7 & 2.479    & 19.274 $\pm$ 0.011   & 137.96108613064 & 24.09605291507 & 19.799 $\pm$ 0.010     & 19.378 $\pm$ 0.042     & 18.764 $\pm$ 0.050      \\
                   &     &   & 137.96091707746 & 24.09603443454 & 19.892 $\pm$ 0.014    & 19.498 $\pm$ 0.467     & 18.738 $\pm$ 0.175     \\
100229.47+444942.8 & 2.048    & 18.425 $\pm$ 0.007   & 150.62297307375 & 44.82851082314 & 18.940  $\pm$ 0.010     & 18.511 $\pm$ 0.038     & 17.888 $\pm$ 0.028     \\
                   &     &   & 150.62269763487 & 44.82856564947 & 19.012 $\pm$ 0.006    & 18.519 $\pm$ 0.030      & 17.903 $\pm$ 0.032     \\
122518.66+483116.3 & 3.082    & 18.131 $\pm$ 0.006   & 186.32777588616 & 48.5211465442  & 18.008 $\pm$ 0.004    & 18.100   $\pm$ 0.014     & 17.253 $\pm$ 0.012     \\
                   &     &    & 186.32778403906 & 48.5213967025  & 19.258 $\pm$ 0.004    & 18.151 $\pm$ 0.135     & 17.338 $\pm$ 0.105     \\
161320.01+170839.2 & 1.547    & 18.889 $\pm$ 0.009   & 243.33351087171 & 17.14416276643 & 19.453 $\pm$ 0.005    & 19.021 $\pm$ 0.093     & 18.506 $\pm$ 0.071     \\
                   &    &   & 243.3333279996  & 17.14427796005 & 19.486 $\pm$ 0.011    & 18.907 $\pm$ 0.050      & 18.468 $\pm$ 0.032\\
\hline
\end{tabular}
\label{table:tb2}
\end{table*}

\begin{table*}
\caption{Basic properties of the LQ candidates from sub-sample B. The known LQs and the reported LQ candidates have been labeled with stars and pluses in the upper right of their source ids, respectively.}
\centering
\begin{tabular}{lllllllll}
\hline  \hline
\textbf{name} &  \textbf{redshift} & \textbf{SDSS $g$} & \textbf{ra$_\textrm{Gaia}$} & \textbf{dec$_\textrm{Gaia}$} & \textbf{$G$} & \textbf{$\rm G_\textrm{BP}$} &\textbf{$\rm G_\textrm{RP}$} \\ \hline
081331.28+254503.0$^{*}$ & 1.512    & 16.092 $\pm$ 0.003   & 123.38030149239 & 25.75086321438 & 16.338 $\pm$ 0.012    & 16.211 $\pm$ 0.010      & 15.174 $\pm$ 0.008     \\
                   &        &         & 123.38054144472 & 25.75079410426 & 17.940 $\pm$ 0.007    &      &         \\
091127.61+055054.1$^{*}$ & 2.798    & 18.028 $\pm$ 0.006   & 137.86513574363  & 5.84841020213    & 18.870  $\pm$ 0.014    & 18.397 $\pm$ 0.028     & 17.762 $\pm$ 0.016      \\
                   &                 &         & 137.86423192750 & 5.848517226935   & 19.875 $\pm$ 0.008    & 20.073 $\pm$ 0.092     & 19.423 $\pm$ 0.111      \\
                   &                &         & 137.86505887342  & 5.84829857367    & 19.508 $\pm$ 0.020       &  &    \\
                    &                &         & 137.86505738812 & 5.848563639185 & 19.745 $\pm$ 0.016       &     &           \\
163348.98+313411.9$^{*}$ & 1.523    & 17.371 $\pm$ 0.004   & 248.4540730856   & 31.5700034202    & 17.432 $\pm$ 0.004    & 17.432 $\pm$ 0.011     & 16.795 $\pm$ 0.010         \\
                   &                &         & 248.45424348917  & 31.56989117914   & 19.132 $\pm$ 0.006    &        &          \\
095122.57+263513.9$^{*}$ & 1.249    & 17.729 $\pm$ 0.005   & 147.84403947441  & 26.58723286072   & 17.490  $\pm$ 0.006    & 17.522 $\pm$ 0.019     & 16.855 $\pm$ 0.015      \\
                   &                 &         & 147.84431797953  & 26.5870561662    & 19.199 $\pm$ 0.009    &        &                 \\
102111.01+491330.3$^{*}$ & 1.721    & 19.431 $\pm$ 0.011   & 155.29599915159  & 49.22503654914   & 19.525 $\pm$ 0.009    & 19.423 $\pm$ 0.036     & 18.522 $\pm$ 0.026     \\
                   &                 &         & 155.2956378427   & 49.22525313164   & 20.998 $\pm$ 0.025    &        &                        \\
112818.49+240217.4$^{*}$ & 1.608    & 18.295 $\pm$ 0.007   & 172.07698911386 & 24.03815267311 & 18.531 $\pm$ 0.008    & 18.383 $\pm$ 0.034     & 17.598 $\pm$ 0.031     \\
                   &              &         & 172.07717757309 & 24.03828375358 & 19.317 $\pm$ 0.008    &         &      \\
234330.58+043557.9$^{*}$ & 1.605    & 18.732 $\pm$ 0.006   & 355.87759501346 & 4.5993938917   & 18.635 $\pm$ 0.011    & 18.654 $\pm$ 0.046 & 17.489 $\pm$ 0.034    \\
                   &          &         & 355.87727347014 & 4.59951311798  & 18.961 $\pm$ 0.008    &     &         \\                   
105926.43+062227.4$^{+}$ & 2.199    & 17.633 $\pm$ 0.005   & 164.86015095584 & 6.37420498672  & 17.462 $\pm$ 0.015    & 17.205 $\pm$ 0.024     & 16.538 $\pm$ 0.017   \\
                   &              &         & 164.86009996452 & 6.37436131019  & 17.946 $\pm$ 0.020     &      &             \\
132128.67+541855.5$^{+}$  & 2.259    & 19.974 $\pm$ 0.019   & 200.36940900851 & 54.31544057296 & 20.265 $\pm$ 0.013    & 20.185 $\pm$ 0.097     & 19.477 $\pm$ 0.070      \\
                   &              &         & 200.36978382582 & 54.31540192645 & 20.943 $\pm$ 0.033    &         &     \\
212243.01-002653.6$^{+}$  & 1.975    & 19.025 $\pm$ 0.009   & 320.67921837391  & -0.44828161549   & 19.270  $\pm$ 0.006    & 19.078 $\pm$ 0.043     & 18.243 $\pm$ 0.038          \\
                   &                &         & 320.67912428067  & -0.44816230327   & 19.860  $\pm$ 0.009    &        &               \\
205752.47+000635.2$^{+}$ & 1.135    & 18.943 $\pm$ 0.009   & 314.46863940567 & 0.10978924318  & 18.718 $\pm$ 0.004    & 18.744 $\pm$ 0.016     & 18.201 $\pm$ 0.031      \\
                   &                  &         & 314.46874170038 & 0.11004439111  & 20.407 $\pm$ 0.007    &      &         \\
162501.98+430931.6$^{+}$ & 1.657    & 18.978 $\pm$ 0.009   & 246.25829345634 & 43.15872055958 & 19.233 $\pm$ 0.011    & 19.007 $\pm$ 0.034     & 18.369 $\pm$ 0.042    \\
                   &              &         & 246.25818775771 & 43.15885994875 & 20.016 $\pm$ 0.010     &        &      \\
003715.87+205825.6 & 2.049    & 19.658 $\pm$ 0.017   & 9.31620407013   & 20.9737723715  & 19.458 $\pm$ 0.011    & 19.358 $\pm$ 0.061     & 18.617 $\pm$ 0.037     \\
                   &          &                 & 9.31590244232   & 20.97379245651 & 20.460  $\pm$ 0.020     &                    &           \\
014613.39+291548.7 & 1.095    & 18.385 $\pm$ 0.007   & 26.55577686969  & 29.26353639033 & 18.437 $\pm$ 0.006    & 18.571 $\pm$ 0.026     & 17.776 $\pm$ 0.026     \\
                   &          &               & 26.55596783268  & 29.2636809947  & 20.349 $\pm$ 0.014    &              &           \\
074942.51+171512.1 & 2.165    & 19.212 $\pm$ 0.011   & 117.42710879551 & 17.25336133864 & 19.017 $\pm$ 0.007    & 19.070  $\pm$ 0.031     & 18.266 $\pm$ 0.031     \\
                   &          &                 & 117.42738076567 & 17.25340257501 & 20.500   $\pm$ 0.012    &               &           \\
082341.07+241805.4 & 1.814    & 17.490  $\pm$ 0.005   & 125.92116891379 & 24.30157031702 & 17.915 $\pm$ 0.006    & 17.711 $\pm$ 0.020      & 16.721 $\pm$ 0.022     \\
                   &          &                 & 125.92115552747 & 24.30139480146 & 18.253 $\pm$ 0.007    &               &           \\
094046.85+471421.1 & 1.394    & 19.419 $\pm$ 0.012   & 145.19521322175 & 47.23919254429 & 19.475 $\pm$ 0.007    & 19.376 $\pm$ 0.048     & 18.622 $\pm$ 0.029     \\
                   &          &                 & 145.19524563209 & 47.23939009799 & 20.715 $\pm$ 0.013    &               &           \\
093801.86+421603.3 & 1.708    & 19.059 $\pm$ 0.010    & 144.50770088391 & 42.26761835091 & 19.723 $\pm$ 0.009    & 19.410  $\pm$ 0.044     & 18.770  $\pm$ 0.032     \\
                   &          &                & 144.50793689869 & 42.26748578136 & 20.055 $\pm$ 0.009    &                &           \\
091704.23+312810.7 & 1.168    & 17.801 $\pm$ 0.005   & 139.26763531237 & 31.46963605502 & 17.682 $\pm$ 0.009    & 17.759 $\pm$ 0.028     & 17.242 $\pm$ 0.022     \\
                   &          &                 & 139.26738611027 & 31.46968976471 & 19.894 $\pm$ 0.009    &               &           \\
103451.36+361331.4 & 0.510     & 18.312 $\pm$ 0.007   & 158.71403496132 & 36.22533464576 & 18.774 $\pm$ 0.008    & 18.575 $\pm$ 0.034     & 17.987 $\pm$ 0.040      \\
                   &          &                 & 158.71406359898 & 36.22557455555 & 20.374 $\pm$ 0.013    &                &           \\
101504.76+123022.2 & 1.705    & 17.625 $\pm$ 0.005   & 153.76982886867 & 12.50619163833 & 17.359 $\pm$ 0.006    & 17.455 $\pm$ 0.015     & 16.92 $\pm$ 0.009     \\
                   &          &                 & 153.76965447866 & 12.50611075649 & 19.716 $\pm$ 0.014    &               &           \\
101329.41+443829.8 & 3.011    & 19.132 $\pm$ 0.011   & 153.37259938695 & 44.64159607963 & 18.931 $\pm$ 0.005    & 19.098 $\pm$ 0.025     & 18.356 $\pm$ 0.023     \\
                   &          &                & 153.3723606294  & 44.64181558952 & 20.937 $\pm$ 0.035    &               &           \\
095636.42+690028.3 & 1.975    & 18.266 $\pm$ 0.007   & 149.15203897788 & 69.0078790747  & 18.165 $\pm$ 0.004    & 18.161 $\pm$ 0.016     & 17.335 $\pm$ 0.017     \\
                   &          &                 & 149.15154197643 & 69.00778738764 & 19.548 $\pm$ 0.009    &                &           \\
111853.50+074946.0 & 2.042    & 18.935 $\pm$ 0.009   & 169.72295455365 & 7.82939200109  & 18.842 $\pm$ 0.006    & 18.768 $\pm$ 0.032     & 18.143 $\pm$ 0.026     \\
                   &          &                 & 169.72292312097 & 7.82953941198  & 19.952 $\pm$ 0.014    &               &           \\
120012.26+160724.5 & 2.137    & 18.804 $\pm$ 0.008   & 180.05111651692 & 16.12351965363 & 18.819 $\pm$ 0.006    & 18.897 $\pm$ 0.027     & 18.314 $\pm$ 0.026     \\
                   &          &                 & 180.05112084092 & 16.1232566556  & 20.842 $\pm$ 0.021    &               &           \\
130505.67-023219.0 & 1.374    & 18.983 $\pm$ 0.010    & 196.27367026925 & -2.53866483065 & 18.903 $\pm$ 0.007    & 18.883 $\pm$ 0.035     & 18.199 $\pm$ 0.046     \\
                   &          &                 & 196.27366189289 & -2.53843594023 & 20.299 $\pm$ 0.011    &                &           \\
125640.55+105742.5 & 0.773    & 18.419 $\pm$ 0.007   & 194.16891486138 & 10.96185898762 & 18.744 $\pm$ 0.006    & 18.638 $\pm$ 0.032     & 18.297 $\pm$ 0.021     \\
                   &          &                 & 194.16912055423 & 10.96190030544 & 20.627 $\pm$ 0.012    &              &           \\
123846.68+644836.6 & 1.560     & 17.968 $\pm$ 0.006   & 189.69441220471 & 64.81015615149 & 18.502 $\pm$ 0.006    & 18.382 $\pm$ 0.021     & 17.509 $\pm$ 0.022     \\
                   &          &                & 189.69506922748 & 64.81025202878 & 21.115 $\pm$ 0.026    &                &     \\
151109.85+335701.6 & 0.799    & 18.707 $\pm$  0.008   & 227.79107461147 & 33.95044311403 & 19.004 $\pm$  0.005    & 18.845 $\pm$  0.031     & 18.282 $\pm$  0.025     \\
                   &          &                & 227.7910201002  & 33.95074072852 & 20.087 $\pm$ 0.006    &              &           \\
\hline  \hline
\label{table:tb3}
\end{tabular}
\end{table*}

\begin{table*}
\centering
\begin{tabular}{lllllllll}
\hline  \hline
142322.93+154437.5 & 1.325    & 19.440  $\pm$ 0.012   & 215.84552422351 & 15.74377030263 & 19.531 $\pm$ 0.01     & 19.477 $\pm$ 0.058     & 18.714 $\pm$ 0.044     \\
                   &          &                 & 215.84575905158 & 15.74379606868 & 20.753 $\pm$ 0.018    &               &           \\
142341.94+204001.5 & 0.842    & 18.970  $\pm$ 0.009   & 215.92477238695 & 20.66712154342 & 19.12  $\pm$ 0.01     & 19.153 $\pm$ 0.041     & 18.559 $\pm$ 0.034     \\
                   &          &                & 215.92463966802 & 20.66685847755 & 20.903 $\pm$ 0.029    &                &           \\
141742.28+243723.4 & 0.692    & 18.739 $\pm$ 0.008   & 214.42621742001 & 24.62319243904 & 18.676 $\pm$ 0.005    & 18.594 $\pm$ 0.021     & 18.113 $\pm$ 0.017     \\
                   &          &                 & 214.42606048912 & 24.62300673577 & 20.523 $\pm$ 0.011    &              &           \\
161153.19+054947.9 & 1.318    & 18.720  $\pm$ 0.008   & 242.97162386769 & 5.82998682859  & 18.994 $\pm$ 0.004    & 19.162 $\pm$ 0.036     & 18.315 $\pm$ 0.021     \\
                   &          &                & 242.97153951361 & 5.82978515317  & 21.097 $\pm$ 0.03     &               &           \\
212315.32+095050.6 & 0.437    & 19.321 $\pm$ 0.013   & 320.81394284429 & 9.84747613285  & 19.729 $\pm$ 0.011    & 19.171 $\pm$ 0.066     & 18.340  $\pm$ 0.035     \\
                   &          &               & 320.81378526283 & 9.84730546774  & 19.994 $\pm$ 0.008    &                &           \\
211731.94+090237.1 & 0.960     & 19.410  $\pm$ 0.012   & 319.38306783267 & 9.04364377762  & 19.343 $\pm$ 0.013    & 19.197 $\pm$ 0.042     & 18.596 $\pm$ 0.038     \\
                   &          &                & 319.38321747176 & 9.04364517363  & 20.193 $\pm$ 0.018    &               &           \\
170256.01+252444.9 & 2.177    & 19.057 $\pm$ 0.011   & 255.73339296378 & 25.41257027385 & 19.233 $\pm$ 0.005    & 19.183 $\pm$ 0.041     & 18.512 $\pm$ 0.028     \\
                   &          &                 & 255.73328397873 & 25.4122801499  & 20.122 $\pm$ 0.007    &              &           \\
223512.90+314611.6 & 2.248    & 19.069 $\pm$ 0.010    & 338.80379947433 & 31.76989519099 & 18.989 $\pm$ 0.004    & 19.136 $\pm$ 0.027     & 18.416$\pm$ 0.021     \\
                   &          &                 & 338.80345519001 & 31.76996483494 & 20.974 $\pm$ 0.017    &               &           \\
222819.72+173255.7 & 2.125    & 18.898 $\pm$ 0.009   & 337.08219088284 & 17.54880244878 & 18.855 $\pm$ 0.008    & 18.949 $\pm$ 0.033     & 18.278 $\pm$ 0.033     \\
                   &          &                 & 337.08196086552 & 17.54895879448 & 20.564 $\pm$ 0.011    &              &           \\
221217.12+035040.5 & 0.214    & 17.411 $\pm$ 0.005   & 333.07131876628 & 3.8445960753   & 18.037 $\pm$ 0.025    & 17.932 $\pm$ 0.078     & 17.130  $\pm$ 0.043     \\
                   &          &                 & 333.0711619343  & 3.84480172068  & 20.785 $\pm$ 0.023    &              &           \\
220531.24+232339.5 & 1.169    & 19.138 $\pm$ 0.010    & 331.38020828903 & 23.39432374363 & 19.009 $\pm$ 0.004    & 19.098 $\pm$ 0.031     & 18.322 $\pm$ 0.023     \\
                   &          &                 & 331.37996426152 & 23.3941929473  & 20.693 $\pm$ 0.011    &               &           \\
233522.51+320109.1 & 0.904    & 19.683 $\pm$ 0.013   & 353.8438295556  & 32.01918772811 & 19.458 $\pm$ 0.005    & 19.370  $\pm$ 0.036     & 18.699 $\pm$ 0.028     \\
                   &          &                 & 353.84384120973 & 32.01935785875 & 20.547 $\pm$ 0.019    &                &           \\
233348.78+285412.8 & 1.671    & 18.407 $\pm$  0.007   & 353.45329231124 & 28.90360063212 & 18.191 $\pm$  0.004    & 18.312 $\pm$  0.019     & 17.618 $\pm$  0.015     \\
                   &          &                 & 353.45328903418 & 28.90338416532 & 20.301 $\pm$ 0.01     &               &   \\ 
\hline  \hline
\end{tabular}
\end{table*}

\begin{figure*}
\centering
\includegraphics[width=1.0\textwidth]{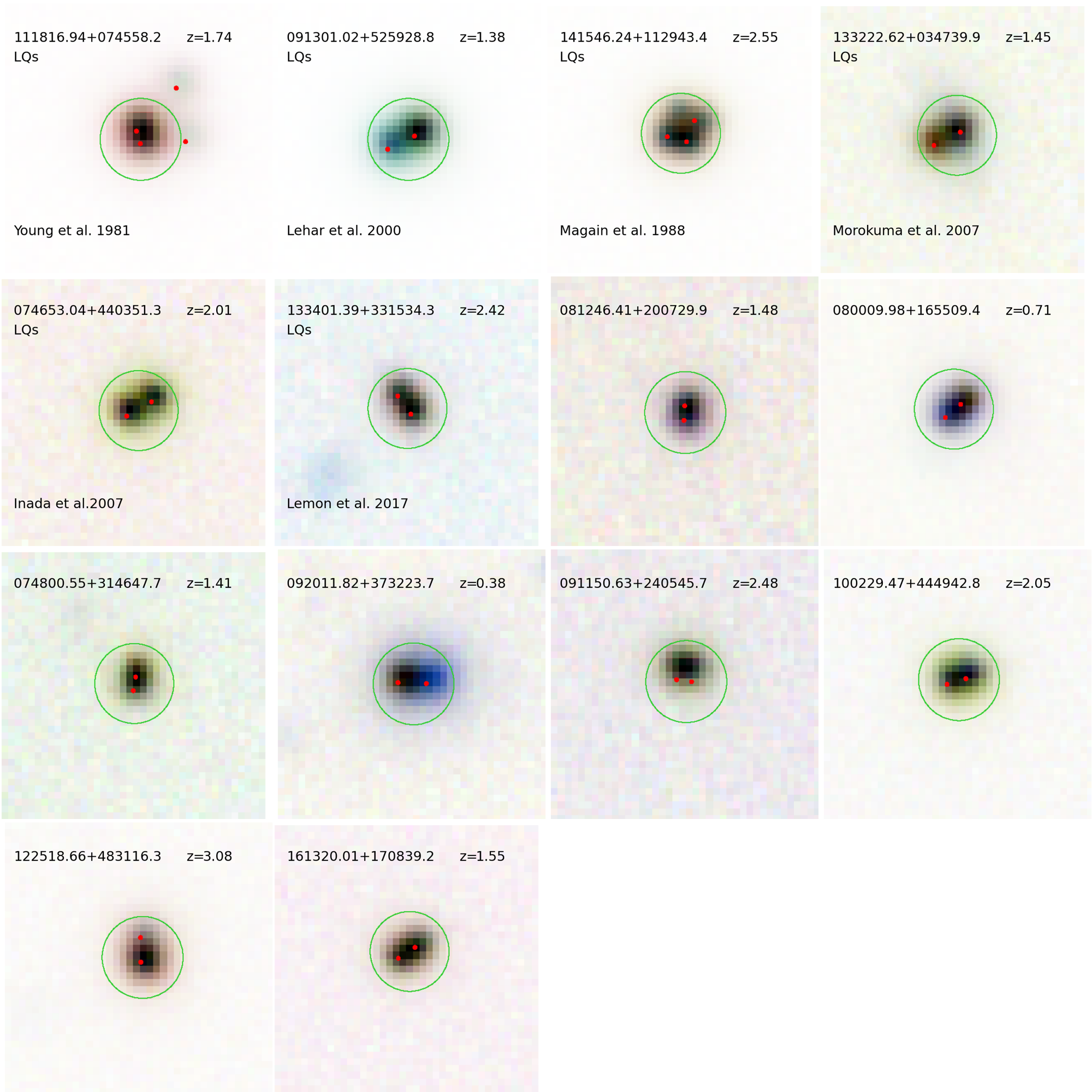}
\caption{Pan-STARRS composite colour images of the selected LQ candidates from the sub-sample A, which are listed in Table \ref{table:tb2}. The legends are similar to that in the right panels of Figure \ref{fig:f4}.
}
\label{fig:f6}
\end{figure*}

\subsection{LQ candidates}
We use the colour difference method to select LQ candidates, which is based on the fact that the colour difference between the multiple lensed images is usually very small for one LQ system. For the remaining 17 targets in sub-sample A, we show the colour difference of $\rm (G_{BP}-G_{RP})_{d1} - (G_{BP}-G_{RP})_{d2}$ as a relationship of the separations as indicated in the left panel of Figure \ref{fig:f5}. The separations are calculated from the coordinates measured from \textit{Gaia} in one system. If the system has multiple \textit{Gaia} detections, we choose the detections with measurements of $\rm G_{BP}$, and $\rm G_{RP}$ bands preferentially to carry out the color difference analysis. Also, we have shown the results of the known LQs, known DQs, and DQ candidates in this sub-sample for comparison. Finally, we set the value of $\pm$ 0.4 of the colour difference to select the LQ candidates in sub-sample A, which is consistent with that in \cite{Delchambre2019}. By applying this criteria, We have selected 14 LQ candidates including all the 6 known ones from sub-sample A.

For the remaining 61 targets in sub-sample B, we use \textit{Gaia} $\rm G_{BP}$ and $\rm G_{RP}$ magnitudes as well as SDSS {\it g} and {\it i} bands to obtain the colour difference as described in section \ref{sec:method}. To set the limit of the selection criteria, we use the targets in sub-sample A as a reference as shown in the middle panel of Figure \ref{fig:f5}. Based on the distributions for the known LQs, we set the following criteria to select the LQ candidates for those targets in sub-sample B, ie, 0.2<$\rm (G_{BP}-G_{RP})_{d1} -(g -i)_{tot}$ <0.6. The right panel of Figure \ref{fig:f5} shows the result of the colour difference for the remaining 61 targets in sub-sample B. Also, we have labeled the reported LQs, LQ candidates, reported DQs, and DQ candidates from previous works for comparison. Via this criteria, we have identified most of the known LQs quasars or candidates (7 out of 8 known LQs and all the reported LQ candidates) in sub-sample B as LQ candidates.

\begin{figure*}
\centering
\includegraphics[width=1.0\textwidth]{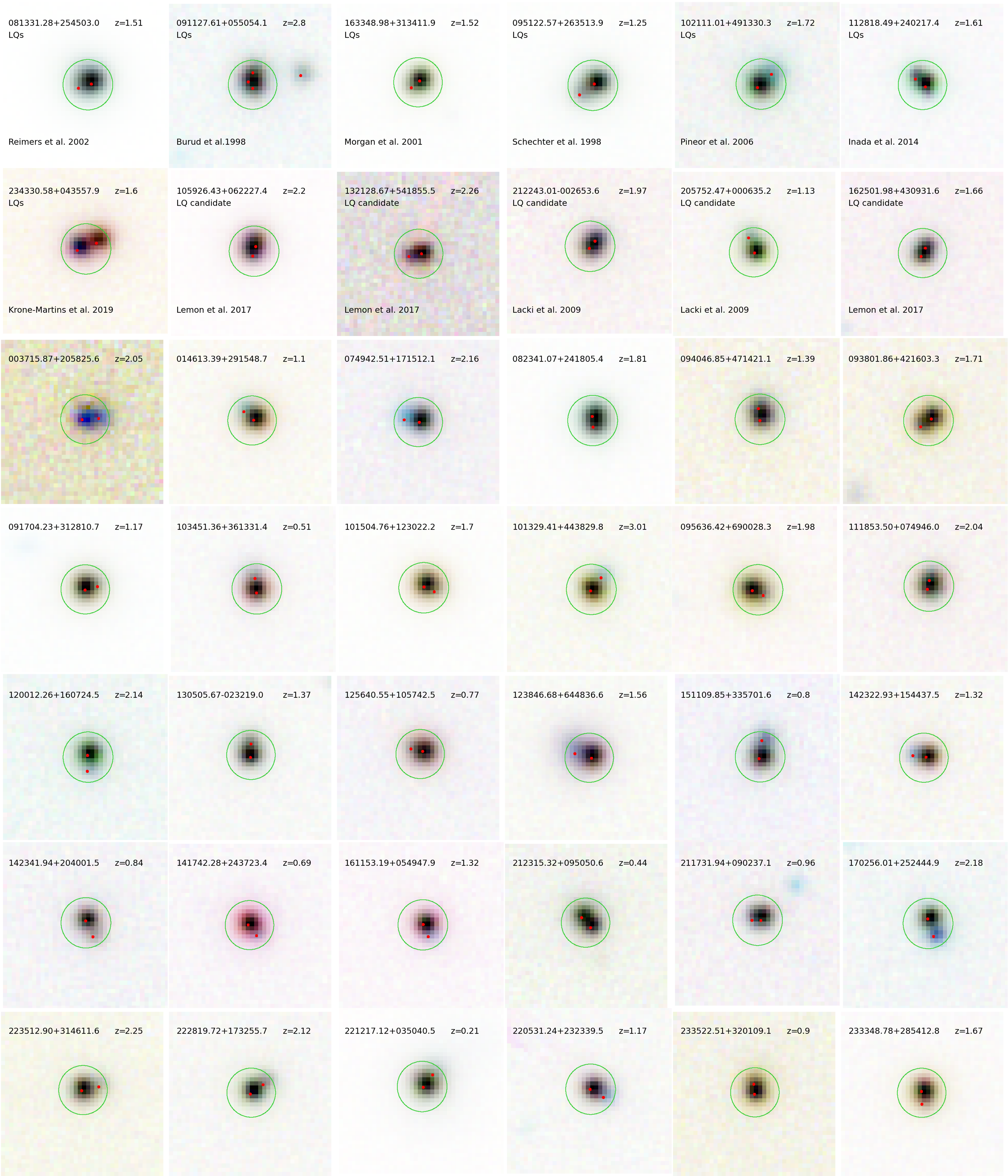}
\caption{Legends similar to that in Figure~\ref{fig:f6} but for the LQ candidates from sub-sample B which are listed in Table~\ref{table:tb3}.
}
\label{fig:f7}
\end{figure*}

Based on the above analysis, we finally select 14 LQ candidates (including 6 known LQs) in sub-sample A and 42 LQ candidates (including 7 known LQs and 5 reported LQ candidates) in sub-sample B. The selected LQ candidates from sub-sample A and B are summarized in Table \ref{table:tb2} and  \ref{table:tb3}, respectively. Furthermore, we show the optical Pan-STARRS composite colour images of the selected LQ candidates in Figure \ref{fig:f6} and Figure \ref{fig:f7}. However, their structures are still elusive, and further high spatial-resolved spectroscopic confirmation is needed to finally pin them down from alternatives.

\subsection{Distributions of the Separations as a Function of Redshift}
\label{subsec:distribution}

\begin{figure*}
\centering
\includegraphics[width=1.0\textwidth]{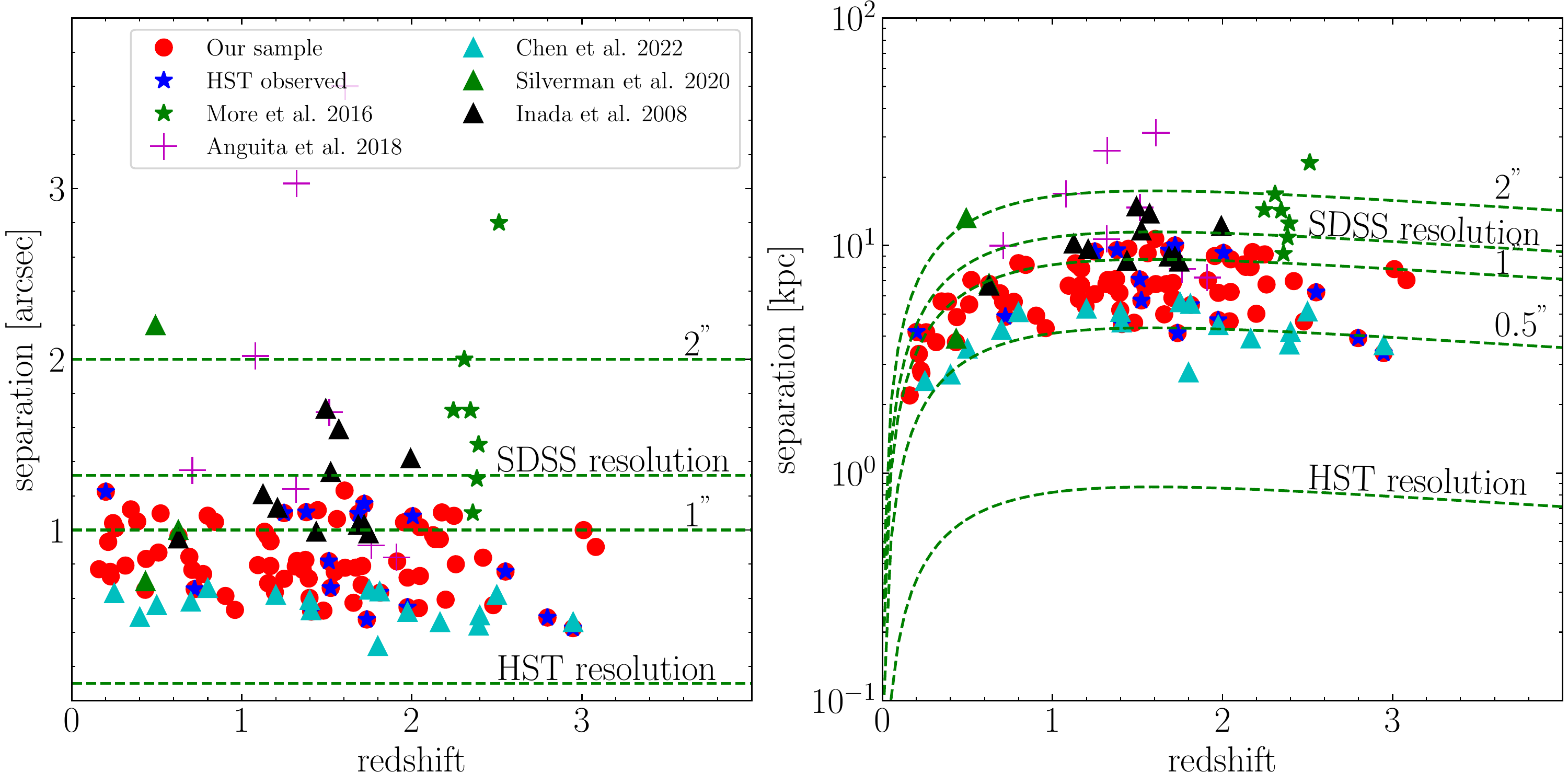}
\caption{Distributions of the redshift verse the angular separation in arcsec (left panel) and the physical separation in kpc scale (right panel) for the targets in our sample, respectively. The red dots represent the targets in our sample and we mark the sources with HST observations in blue stars. For comparison, we also include the DQs or LQs from \citet[][]{Anguita2018}, \citet[][]{Silverman2020}, \citet[][]{Chen2022}, \citet[][]{Inada2008} and \citet[][]{More2016} in purple pluses, green triangles, cyan triangles, black triangles, and green stars, respectively. The resolution of SDSS and HST are marked in green dotted lines for reference. 
}
\label{fig:f8}
\end{figure*}

\begin{figure*}
\centering
\includegraphics[width=1.0 \textwidth]{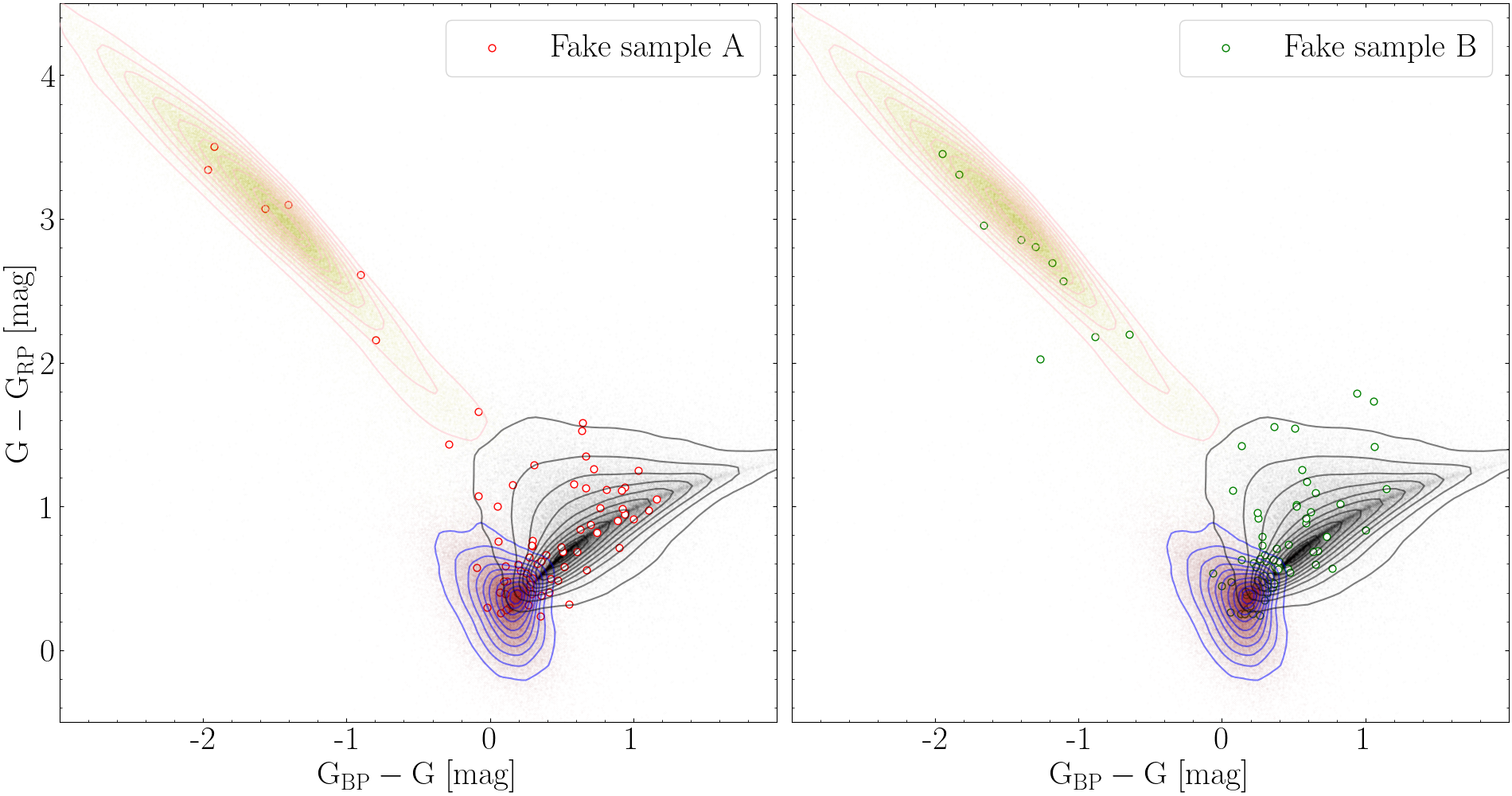}
\caption{
Colour-colour diagram for the \textit{Gaia} detections with measurements of magnitudes of \textit{G}, $\rm G_{BP}$, and $\rm G_{RP}$ passbands for the remaining targets in the fake samples A and B in the left and right panels, respectively. The background is similar to that in Figure \ref{fig:f3}.
}
\label{fig:f9}
\end{figure*} 

Since there are at least two detections from \textit{Gaia} for the sources in our sample when cross-matched with SDSS DR16Q, thus the angular and the physical separations can be calculated accordingly. The distributions of the angular and projected separations for individual remaining targets after Step One as a function of the redshift are shown in the left and right panels of Figure \ref{fig:f8}, respectively. For the sources with more than two \textit{Gaia} detections, we show the smallest separations between the multiple detections from \textit{Gaia} in one system. We also show the resolution of SDSS and HST in blue and green dotted lines in both panels in Figure \ref{fig:f8} for reference.

From Figure \ref{fig:f8}, we could see that the angular separations of the sources are about 1\arcsec\ more or less, corresponding to projected separations in the range from 1 to 10 kpc at the redshift range of z = 0 $\sim$ 3.5. And all the angular or physical separations of the sources in our sample are beyond the limit of the HST resolution, with only the separations of five sources larger than the limit resolution of SDSS. We also show the known DQs or LQs with multiple detections from \textit{Gaia} for comparison. Generally, the separations of our sample are typically larger than those in \citet[][]{Chen2022} but smaller than those in \citet[][]{More2016} and \citet[][]{Inada2008}.

\section{Discussions}
\label{sec:discussion}

\subsection{Discussion of our selection methods}
\label{sec:method_dis}

To improve the accuracy and help us better constrain the types of those abnormal quasars, we have introduced a succession of different methods. Here we discuss briefly about our current classification strategy:

\begin{itemize}
    \item  Stellar features: we have introduced a method to pick out the quasar-star pairs via their stellar features, including the proper motion or parallax criteria, stellar features in the SDSS spectra, and via the colour-colour diagram. We should note that the proper motion or parallax criteria would mis-identify quasar-star pairs. This has been pointed out by \citet{Makarov2022ApJ} that a small portion of quasars with excess proper motions turn out to be either dual quasars or lensed quasars. Furthermore, we consider the effects of quasar-star superposition and estimate the probability in section \ref{sec:superposition}. For the quasar-star pairs selected via the stellar features in the SDSS spectra, they all have been picked out by the proper motion or the parallax criteria. As for the colour-colour diagram proposed by \cite{Bailer-Jones2019}, which uses a machine learning method to classify the objects from \textit{Gaia} DR2 into stars, quasars, and galaxies. The spectroscopic classifications cross-matched with \textit{Gaia} is served as the training sets for their justifications. Hence the colour-colour diagram is based on the statistical distributions of a large sample. However, there are still some uncertainties around the boundaries between the three types, which may lead to bias when applied to our sample. 
    
    \item Emission line features: we have found a promising way to find the DQ candidates by combining the line offset or the double-peaked properties of the emission lines as well as the multiple detections from \textit{Gaia}. However, the double-peaked or offset properties can also be explained by different physical interpretations. Future high-resolution spectra are still needed to pin them down from alternatives.

    \item Colour difference: there are still some uncertainties based on our current colour selection strategy. Firstly, we set the limits of the colour difference according to the known LQs and DQs in our sample. As a consequence, there are still some uncertainties due to the small sample of the confirmed LQs or DQs in our sample. While including the known LQs with larger separations would bring in a mis-match of other nearby objects when carrying out the colour analysis. Secondly, the colour difference may also be contaminated by the lensing galaxies. Thirdly, it's hard to separate the DQs from the LQs without follow-up spectroscopic confirmations. The reported DQ SDSS J084129.77+482548.3 is also classified as LQ candidates via our colour difference method. Since some DQs may also take on similar colors while some not as determined by the redshifts of the quasars. We could not exclude the possibility that the selected LQ candidates maybe turn out to be DQs. We tend to regard those targets with similar color differences as LQ candidates in this work.

\end{itemize}

\subsection{Quasar-star superposition}
\label{sec:superposition}

Although we have excluded the quasar-star pairs by searching for their stellar features, there are still some possibilities that these systems are DQs or LQs with a superpositioned star. We find that two reported LQs, SDSS 081331.28+254503.0 \citep{Reimer2002,Jackson2015} and SDSS J234330.58+043557.9 \citep{Krone-Martins2019}, are classified as quasar-star pairs according to our proper motion/parallax criteria. Both of these two LQs have two \textit{Gaia} detections. We think that \textit{Gaia} only detected one of the lensed images and the other \textit{Gaia} detection came from a superpositioned star with large proper motion or parallax. The HST archival images of the known LQ SDSS 081331.28+254503.0 clearly showed the lensed quasar system and \textit{Gaia} only detected the brightest lensed image and a superpositioned star. The situation is also similar for some DQs. For our selected dual quasar candidate SDSS J221130.31+291949.0, we also find that one \textit{Gaia} detection was observed with large proper motion. However, we still propose it as a dual quasar candidate in this work considering its line offset properties as well as the two detections from \textit{Gaia}. We are carrying up follow-up spectroscopic confirmation to pin it down from alternatives.

We also test the quasar-star superposition by using the method proposed by \citet{Chen2022}. For the SDSS DR16q quasars, we offset their decl. by $+$ or $-$ $1^{\prime}$ and search for nearby \textit{Gaia} DR3 sources at the offset positions within 1" radius. We require that the G-band magnitude $ \leq $ 20.7 mag for the cross-matched targets. We find 655 and 653 targets fulfilling the above conditions, which are labeled as fake sample A and fake sample B, respectively. We use the proper motion/parallax criteria and the colour-colour diagram method in Step one to classify quasar-star pairs in both samples. We find that 584 and 572 targets could be selected by the proper motion/parallax criteria in fake samples A and B, respectively. For the remaining targets, 37 and 41 targets could be further selected by the colour-colour diagram method as shown in the left and right panels of Figure \ref{fig:f9}. Therefore, it indicates that $\sim $ 95 percent of the targets in fake sample A and $\sim $ 94 percent in fake sample B could be selected as quasar-star pairs by the proper motion/parallax criteria and the colour-colour diagram method. As for our sample, it indicates that $\sim$ 4 targets in the remaining 78 targets after Step One are quasar-star pairs.

\subsection{Extra Photometric Errors from Nearby Objects}

As suggested by the referee, here we discuss the reliability of colours for compact double sources. Gaia's $\rm G_{BP}$ and $\rm G_{RP}$ photometric magnitudes are integrated low-resolution spectro-photometry in two different passbands. According to measurement details of low-resolution BP/RP spectra from \textit{Gaia} in section 2.2 from \citet{DeAngeli2022}, only the measurements of the brightest source are recorded and transmitted fully to the ground for the overlapping sources in crowded regions. In other words, if the source in our sample has BP/RP measurements for both detections from \textit{Gaia} in the same system, it should be well-resolved and measured. Otherwise, only the brighter detection has BP/RP detections. Generally, a series of procedures have been developed to deblend the low-resolution spectra in crowded regions for $\rm G_{BP}$ and $\rm G_{RP}$ passbands with \textit{Gaia}. However, the deblending details will not be pubic until the release of \textit{Gaia} DR4.

\begin{figure*}
\centering
\includegraphics[width=1.0\textwidth]{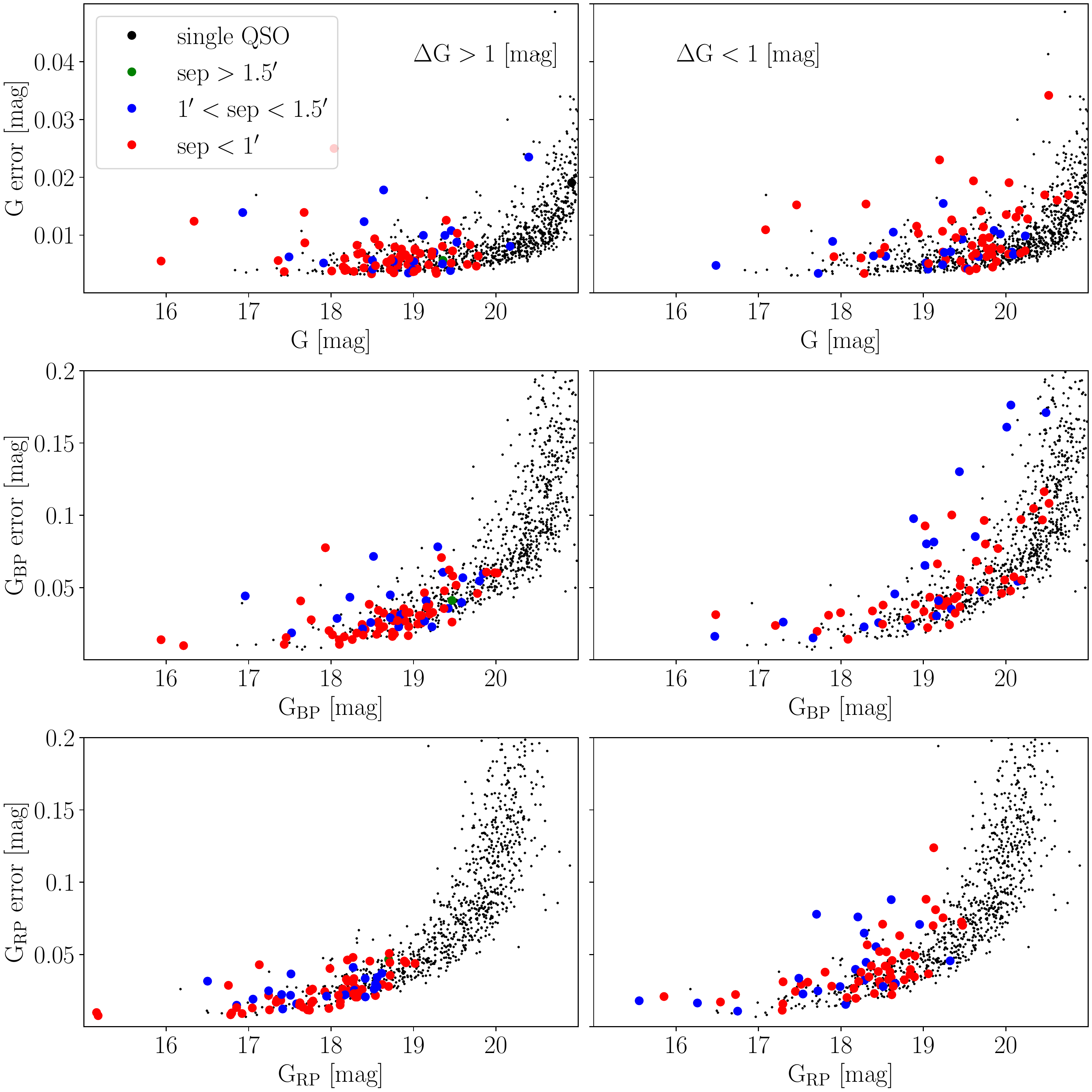}
\caption{ G, $G_{BP}$, and $G_{RP}$ magnitude measurements as a relationship with their errors for the double quasars in our sample. The left and right panels are for the targets which the G band magnitude differences are larger or smaller than 1 mag between the two detections in the same system , respectively. In each separate panel, we mark the targets with the separations larger than 1.5 \arcsec, between 1 \arcsec and 1.5 \arcsec and smaller than 1 \arcsec in green, blue and red dots, respectively. For comparison, we also plot the results from the isolated QSOs from SDSS DR16 cross-matched with \textit{Gaia} with only one \textit{Gaia} detection in small black dots in each panel. 
}
\label{fig:f10}
\end{figure*}

We then use the measurement errors from \textit{Gaia} to judge the extra photometric errors from nearby objects when measuring $\rm G_{BP}$ and $\rm G_{RP}$. We have divided our targets into two groups with the \textit{G} band magnitude differences larger or smaller than 1 mag between the two detections in the same system. We also consider different separations between the two detections in the same system in each group. The results of the photometric measurements as a relationship of measurement errors from \textit{Gaia} for the two groups, as well as isolated Quasars, are shown in Figure \ref{fig:f10}. From this figure, we could see that the dimmer counterpart would certainly bring extra uncertainties when measuring the \textit{G}, $\rm G_{BP}$, and $\rm G_{RP}$ magnitudes, especially for those targets with smaller magnitude differences between the two detections in the same system. The relationship is consistent with that from isolated QSOs as a whole with slightly larger deviations. We also find that there are no distinct differences for the targets with different separations in both sub-samples. Overall, the compact nearby sources would bring in extra errors no more than 0.02, 0.1, and 0.1 for sources brighter than 20 mag in \textit{G}, $\rm G_{BP}$, and $\rm G_{RP}$ bands, respectively.

\subsection{HST coverage}
\begin{figure*}
\centering
\includegraphics[width=0.99 \textwidth]{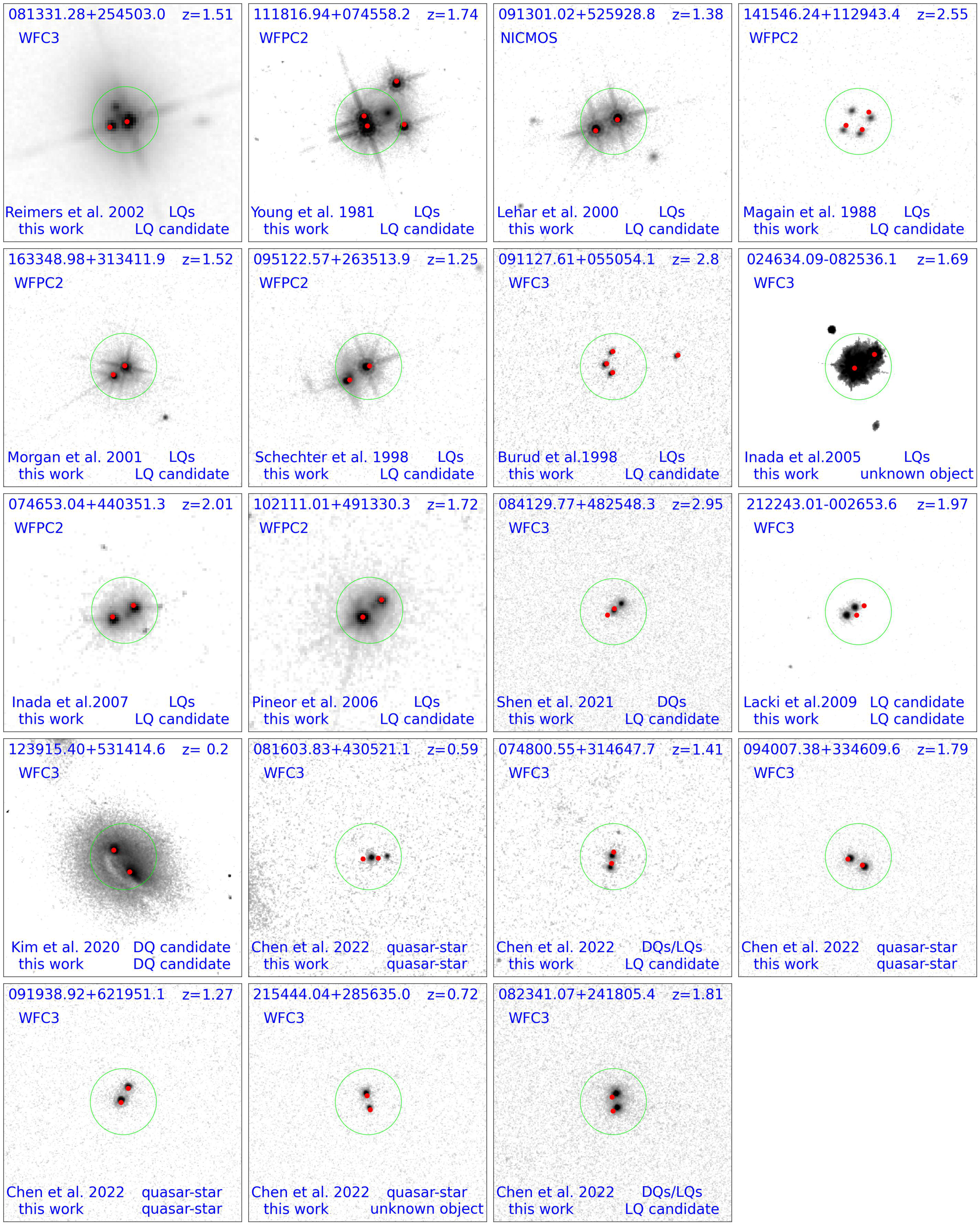}
\caption{Archival HST images of the targets listed in Table \ref{table:tb4}. The resolution for each separate panel is 10\arcsec\  $\times$ 10 \arcsec. In each separate panel, the coordinates from \textit{Gaia} are represented in red points and the green circle indicates a radius of 1.5 \arcsec. The source id, the redshift, the reference, and the classification results are labeled in the top left, top right, bottom left and bottom right, respectively. We also indicate the camera used below the source id in each panel. Previously known LQs, LQ candidates, DQs, and DQ candidates are labeled as 'LQs', 'LQ candidate', 'DQs' and 'DQ candidate', respectively. 
}
\label{fig:f11}
\end{figure*}

Considering the advantages of HST's high-resolution imaging in resolving the physical structures of those strange quasars, we have searched the HST archive and found 19 objects with available HST observations in our sample. The basic properties of all those targets with HST observations are summarised in Table \ref{table:tb4}, with the available archival HST images shown in Figure \ref{fig:f11}. We have also compared our classification results with previous works for those targets in each panel. Our classification results are mainly consistent with previous works, except for the DQ SDSS J084129.77+482548.3. The reason has been discussed in section \ref{sec:method_dis}.

As shown in Figure \ref{fig:f11}, the coordinates measured from \textit{Gaia} match well with that from the HST images for most targets, with only a few of them like SDSS J081603.83+ 430521.1, showing minor offsets. This misalignment may be caused by the coordinates offset from the HST observations. On the other aspect, the observations from HST clearly reveal multiple cores for those objects. This proves that the high-quality astrometric measurement from \textit{Gaia} could be used as a powerful tool when searching for the LQs or DQs.  There are 8 targets without spectroscopic confirmations and we are planning the follow-up spectroscopic observations to reveal their nature.

\subsection{Comparison with related works}
In this work, we mainly focus on the classification of the spectroscopic quasars identified from SDSS DR16q with multiple \textit{Gaia} detections within a radius of 1\arcsec\  from SDSS positions. We introduce our classification methods with the help of the known LQs and DQs in our sample as a reference. There are only a small portion of targets with available HST observations and our classification results for those targets are mainly consistent with previous works.

As described in the introduction, there are also two different methods searching for DQs/LQs with the \textit{Gaia} data. The method of varstrometry has been proved to be very efficient for DQs/LQs with substantial variability. Also, they have found the efficiency of ~53 \% and ~22 \% for both spectroscopic and non-spectroscopic targets \citep{Chen2022}. As for the method of GMP, they show the advantages of selecting DQs/LQs at the separations of 0.1\arcsec to 0.7 \arcsec. The above two methods are applicable for both Gaia-resolved and Gaia-unresolved targets. After cross-matched, we find 8 sources in common with \cite{Shen2021} and \cite{Chen2022}, and 4 sources in common with \citet[][]{Mannucci2022}. Although all the selection methods could have different selection functions, all these prove that the high-resolution measurements from \textit{Gaia} could be an efficient method in searching for DQs or LQs when cross-matched with available quasar catalogs.

\begin{landscape}
\begin{table}
\caption{Basic properties of the targets in our sample with HST observations.}
\label{table:tb4}
    \begin{tabular*}{1.3\textwidth}{@{\extracolsep{\fill}} ccccccccccc}
    \hline
    \textbf{name} & \textbf{redshift} & \textbf{SDSS g} & \textbf{ra$_\textrm{Gaia}$} & \textbf{dec$_\textrm{Gaia}$} &  \textbf{G} & \textbf{$\rm G_{BP}$ } & \textbf{$\rm G_{RP}$} &
    \textbf{camera} & \textbf{class} & \textbf{ref}\\ 
    \hline \hline
081331.28+254503.0 & 1.512    & 16.092 $\pm$ 0.003   & 123.38030149239  & 25.75086321438   & 16.338 $\pm$ 0.012  & 16.211 $\pm$ 0.010    & 15.174 $\pm$ 0.008     & WFC3   & LQs    & {[}1{]}{[}2{]}    \\
                   &                  &         & 123.38054144472  & 25.75079410426   & 17.940  $\pm$ 0.007    &        &                     &        &        &                  \\
111816.94+074558.2 & 1.736    & 16.252 $\pm$ 0.004   & 169.57066749217  & 7.7662504694     & 17.086 $\pm$ 0.011    & 16.481 $\pm$ 0.031     & 15.853 $\pm$ 0.021     & WFPC2  & LQs    & {[}3{]}{[}4{]}   \\
                   &                 &         & 169.57062595971  & 7.76612498888    & 17.211 $\pm$ 0.009    & 16.611 $\pm$ 0.043     & 15.966 $\pm$ 0.033     &        &        &                  \\
                   &                  &         & 169.57025416162 & 7.7666884210447 & 18.528 $\pm$ 0.006    & 18.574 $\pm$ 0.074     & 17.979 $\pm$ 0.096     &        &        &                  \\
                   &                 &         & 169.570157868845 & 7.7661434790373 & 18.866 $\pm$ 0.008    &        &                      &        &        &                  \\
091301.02+525928.8 & 1.380     & 16.868 $\pm$ 0.004   & 138.25417691679  & 52.99138630842   & 16.484 $\pm$0.005    & 16.470  $\pm$ 0.016     & 15.552 $\pm$ 0.018     & NICMOS & LQs    & {[}5{]}          \\
                   &                  &         & 138.25463214963  & 52.99124835768   & 17.068 $\pm$ 0.005    & 16.687 $\pm$ 0.270      & 15.753 $\pm$ 0.294     &        &        &                  \\
141546.24+112943.4 & 2.551    & 17.293 $\pm$ 0.004   & 213.94261159532  & 11.49531201636   & 17.569 $\pm$ 0.015    & 16.788 $\pm$ 0.037     & 16.022 $\pm$ 0.025     & WFPC2  & LQs    & {[}6{]}          \\
                   &                 &         & 213.94281909923  & 11.49536505201   & 17.574 $\pm$ 0.016    & 17.002 $\pm$ 0.109     & 16.057 $\pm$ 0.114     &        &        &                  \\
                   &                 &         & 213.9425258972 & 11.4955317232 & 17.638 $\pm$ 0.019    &        &                     &        &        &                  \\
163348.98+313411.9 & 1.523    & 17.371 $\pm$ 0.004   & 248.4540730856   & 31.5700034202    & 17.432 $\pm$ 0.004    & 17.432 $\pm$ 0.011     & 16.795 $\pm$ 0.010      & WFPC2  & LQs    & {[}7{]}          \\
                   &                &         & 248.45424348917  & 31.56989117914   & 19.132 $\pm$ 0.006    &        &                    &        &        &                  \\
095122.57+263513.9 & 1.249    & 17.729 $\pm$ 0.005   & 147.84403947441  & 26.58723286072   & 17.49  $\pm$ 0.006    & 17.522 $\pm$ 0.019     & 16.855 $\pm$0.015     & WFPC2  & LQs    & {[}8{]}          \\
                   &                 &         & 147.84431797953  & 26.5870561662    & 19.199 $\pm$ 0.009    &        &                    &        &        &                  \\
091127.61+055054.1 & 2.798    & 18.028 $\pm$ 0.006   & 137.86513574363  & 5.84841020213    & 18.870  $\pm$ 0.014    & 18.397 $\pm$ 0.028     & 17.762 $\pm$ 0.016     & WFC3   & LQs    & {[}9{]}          \\
                   &                 &         & 137.86505887342  & 5.84829857367    & 19.508 $\pm$ 0.020     &        &                      &        &        &                  \\
                    &                &         & 137.86505738812 & 5.848563639185 & 19.745 $\pm$ 0.016    &        &                    &        &        &                  \\
                   &                 &         & 137.86423192750 & 5.848517226935   & 19.875 $\pm$ 0.008    & 20.073 $\pm$ 0.092     & 19.423 $\pm$ 0.111     &        &        &                  \\

024634.09-082536.1 & 1.686    & 18.162 $\pm$ 0.006   & 41.64211559961   & -8.42672188068   & 16.929 $\pm$ 0.014    & 16.959$\pm$ 0.044     & 16.502 $\pm$ 0.032     & WFC3   & LQs    & {[}10{]}         \\
                   &                 &         & 41.64186218579   & -8.42654907382   & 18.861 $\pm$ 0.011    &        &                    &        &        &                  \\
074653.04+440351.3 & 2.006    & 19.102 $\pm$ 0.008   & 116.72118927293  & 44.06422019002   & 19.472 $\pm$ 0.009    & 19.019 $\pm$ 0.065     & 18.287 $\pm$ 0.032     & WFPC2  & LQs    & {[}11{]}         \\
                   &                 &         & 116.72082301636  & 44.06436462402   & 19.624 $\pm$ 0.010     & 19.203 $\pm$ 0.098     & 18.369 $\pm$ 0.043     &        &        &                  \\
102111.01+491330.3 & 1.721    & 19.431 $\pm$ 0.011   & 155.29599915159  & 49.22503654914   & 19.525 $\pm$ 0.009    & 19.423 $\pm$ 0.036     & 18.522 $\pm$ 0.026     & WFPC2  & LQs    & {[}12{]}         \\
                   &                 &         & 155.2956378427   & 49.22525313164   & 20.998 $\pm$ 0.025    &        &                     &        &        &                  \\

084129.77+482548.3 & 2.948    & 19.283 $\pm$ 0.010    & 130.37404617611  & 48.43013479612   & 19.478 $\pm$ 0.011    & 19.19 $\pm$ 0.038     & 18.454 $\pm$ 0.038     & WFC3   & DQs    & {[}13{]}{[}14{]} \\
                   &                &         & 130.37417633919  & 48.43005444844   & 19.780  $\pm$ 0.037    & 19.345 $\pm$ 0.091     & 18.267 $\pm$ 0.115     &        &        &                  \\
212243.01-002653.6 & 1.975    & 19.025 $\pm$ 0.009   & 320.67921837391  & -0.44828161549   & 19.27  $\pm$ 0.006    & 19.078 $\pm$ 0.043     & 18.243 $\pm$ 0.038     & WFC3   & LQ candidate & {[}15{]}         \\
                   &                &         & 320.67912428067  & -0.44816230327   & 19.86  $\pm$ 0.009    &        &                   &        &        &                  \\
123915.40+531414.6 & 0.200      & 19.206 $\pm$ 0.007   & 189.8143793831   & 53.2374973296    & 20.399 $\pm$ 0.023    & 18.719 $\pm$ 0.045     & 17.053 $\pm$ 0.019     & WFC3   & DQ candidate & {[}16{]}         \\
                   &                 &         & 189.81404358851  & 53.23722292116   & 21.414 $\pm$ 0.041    & 18.615 $\pm$0.090      & 17.018 $\pm$ 0.048     &        &        &                  \\
081603.83+430521.1 & 0.587    & 20.693 $\pm$ 0.017   & 124.01580245472  & 43.08921103297   & 20.624 $\pm$ 0.016    & 21.003 $\pm$ 0.212     & 19.148 $\pm$ 0.081     & WFC3   &        &                  \\
                   &                 &         & 124.01606562402  & 43.08920154678   & 20.823 $\pm$ 0.034    & 20.193 $\pm$0.013     & 18.975 $\pm$ 0.061     &        &        &                  \\
074800.55+314647.7 & 1.408    & 19.985 $\pm$ 0.015   & 117.00230747636  & 31.77985563261   & 20.033 $\pm$ 0.007    & 19.797 $\pm$ 0.062     & 18.811 $\pm$ 0.051     & WFC3   &        &                  \\
                   &                 &         & 117.00227692679  & 31.77999820592   & 20.319 $\pm$ 0.014    & 19.864 $\pm$ 0.190      & 18.833 $\pm$ 0.041     &        &        &                  \\
094007.38+334609.6 & 1.786    & 19.849 $\pm$ 0.014   & 145.0309042995   & 33.76931645031   & 19.591 $\pm$ 0.008    & 19.446 $\pm$ 0.037     & 18.215 $\pm$ 0.031     & WFC3   &        &                  \\
                   &                  &         & 145.03068493606  & 33.76923979991   & 19.711 $\pm$ 0.006    & 19.451 $\pm$ 0.119     & 18.316 $\pm$ 0.053     &        &        &                  \\
091938.92+621951.1 & 1.270     & 19.378 $\pm$ 0.011   & 139.9122152145   & 62.33088639396   & 18.763 $\pm$ 0.003    & 18.937 $\pm$ 0.017     & 17.974 $\pm$ 0.013     & WFC3   &        &                  \\
                   &                &         & 139.91201908388  & 62.33106405635   & 20.494 $\pm$ 0.008    &      &             &        &        &                  \\
215444.04+285635.0 & 0.723    & 20.002 $\pm$ 0.017   & 328.68353178311  & 28.94315441283   & 19.608 $\pm$ 0.019    & 19.288 $\pm$ 0.040      & 18.712 $\pm$ 0.063     & WFC3   &        &                  \\
                   &                 &         & 328.68348587194  & 28.94297732821   & 20.348 $\pm$ 0.026    & 18.573 $\pm$ 1.295     & 18.776 $\pm$ 0.156     &        &        &                  \\
082341.07+241805.4 & 1.814    & 17.602 $\pm$ 0.005   & 125.92116891379  & 24.30157031702   & 17.915 $\pm$ 0.006    & 17.711 $\pm$ 0.020      & 16.721 $\pm$ 0.022     & WFC3   &        &                  \\
                   &                 &         & 125.92115552747  & 24.30139480146   & 18.253 $\pm$ 0.007    &       &               &        &        & \\
    \hline \hline
    \end{tabular*} 

    \begin{tablenotes}
    \footnotesize
    \item References:
    [1]\cite{Reimer2002};   [2]\cite{Jackson2015};
    [3] \cite{Young1981};  [4] \cite{Chiba2005};
    [5] \cite{Lehar2006};  [6] \cite{Magain1988};
    [7] \cite{Morgan2001};  [8] \cite{Schechter1998}; 
    [9] \cite{Burud1998};  [10] \cite{Inada2005};
    [11] \cite{Inada2007}; [12] \cite{Pindor2006};
    [13]\cite{Shen2021};    [14]\cite{Mannucci2022};  
    [15] \cite{Lacki2009}; [16]\cite{Smith2010}
    \end{tablenotes}
\end{table}
\end{landscape}

\clearpage

\section{Conclusions}
\label{sec:conclusion}

In this paper, we present the optical confirmations of the 143 strange quasars with multiple detections from the \textit{Gaia} EDR3. Our sample mainly consists of spectroscopically identified quasars from SDSS DR16 which have multiple detections from \textit{Gaia} within a radius of 1\arcsec. The angular separations of the targets are $\sim $ 1 arcsec more or less, corresponding to physical separations of several kpc scales. 

Based on the stellar features including proper motion/parallax, stellar-like spectra from SDSS, and colour-colour diagram, we find 65 quasar-star pairs. We also discover 2 DQ candidates, one of which presents a double-peaked [O III] emission line feature and the other shows an obvious broad $H_{\beta}$ velocity offset relative to the [O III] $\lambda$ 5007 line. In addition, we further find 56 LQ candidates with similar colour differences, including 13 known LQs and 5 reported LQ candidates. Although, the resolution of the available Pan-STARRS images is not high enough. With more high-resolution observations in the future, we would certainly reveal their structures and clarify their basic properties. We also check the HST archive for available high-resolution observations and find 19 targets with archival HST observations, of which 10 are reported LQs and 1 reported DQ. Our classification results for those 19 targets are mainly consistent with previous works.

We also see the advantage of the higher resolution of HST compared with Pan-STARRS in resolving the basic structure of those strange quasars. Similarly, the Multiple Channel Imager (MCI) on board the Chinese Space Station Telescope (CSST) has many desirable features including high resolution, deep field of view, and multiple filters. With its broad wavelength coverage and high angular resolution, CSST offers scientific opportunities and a great legacy value that complement other forthcoming space-based and ground-based surveys. CSST is ideally suited for astrometric studies of objects fainter than 20 mag, it will extend Gaia’s mapping of the sky to fainter magnitudes, adding precise positions, parallaxes, distance estimates, and extinction estimates. With the operation of CSST in the future, we will endeavor to advance our research and identify more DQs and LQs. 

\section*{Acknowledgments}
We would like to thank the referee for very helpful comments, especially the suggestion to check the reliability of colours for compact double sources. Z.Y.Z. acknowledges the supports by National Key R\&D Program of China No.2022YFF0503402 and the National Science Foundation of China (12022303). We acknowledge the science research grants from the China Manned Space Project with No.CMS-CSST-2021-A07, No.CMS-CSST-2021-A04, No.CMS-CSST-2021-B04, NO.CMS-CSST-2021-A12 and NO.CMS-CSST-2021-B10. This work has also been supported by the Youth Innovation Promotion Association CAS, the grants from the Natural Science Foundation of Shanghai through grant 21ZR1474100, and National Natural Science Foundation of China (NSFC) through grants 12173069, and 11703065. 

\section*{Data Availability} 
The data sets underlying this article are publicly available. They can be accessed from the \textit{Gaia} archive (https://gea.esac.esa.int/archive/) and the SDSS SkyServer (https://skyserver.sdss.org/dr16/en/tools/search/). The Archival HST images are available via MAST (https://mast.stsci.edu/portal/Mashup/Clients/Mast/Portal.html).

\label{lastpage}

\begin{thebibliography}{99}
%
\bibitem[Agnello et al.(2018)]{Agnello2018} Agnello, A., Schechter, P.~L., Morgan, N.~D., et al.\ 2018, \mnras, 475, 2086. doi:10.1093/mnras/stx3226
%
\bibitem[An et al.(2018)]{An2018} An, T., Mohan, P., \& Frey, S.\ 2018, Radio Science, 53, 1211. doi:10.1029/2018RS006647
%
\bibitem[Anguita et al.(2018)]{Anguita2018} Anguita, T., Schechter, P.~L., Kuropatkin, N., et al.\ 2018, \mnras, 480, 5017. doi:10.1093/mnras/sty2172

\bibitem[Bailer-Jones et al.(2019)]{Bailer-Jones2019} Bailer-Jones, C.~A.~L., Fouesneau, M., \& Andrae, R.\ 2019, \mnras, 490, 5615. doi:10.1093/mnras/stz2947
%
\bibitem[Begelman et al.(1980)]{Begelman1980} Begelman, M.~C., Blandford, R.~D., \& Rees, M.~J.\ 1980, \nat, 287, 307
%
\bibitem[Blackburne et al.(2011)]{Blackburne2011} Blackburne, J.~A., Pooley, D., Rappaport, S., et al.\ 2011, \apj, 729, 34. doi:10.1088/0004-637X/729/1/34

\bibitem[Blackburne et al.(2015)]{Blackburne2015} Blackburne, J.~A., Kochanek, C.~S., Chen, B., et al.\ 2015, \apj, 798, 95. doi:10.1088/0004-637X/798/2/95
%
\bibitem[Burke-Spolaor(2011)]{Burke-Spolaor2011} Burke-Spolaor, S.\ 2011, \mnras, 410, 2113
%
\bibitem[Burud et al.(1998)]{Burud1998} Burud, I., Courbin, F., Lidman, C., et al.\ 1998, \apjl, 501, L5. doi:10.1086/311450
%
\bibitem[Charisi et al.(2016)]{Charisi2016} Charisi, M., Bartos, I., Haiman, Z., et al.\ 2016, \mnras, 463, 2145
%
\bibitem[Chen, Yu \& Lu(2020)]{Chen2020} Chen, Y., Yu, Q., \& Lu, Y.\ 2020, \apj, 897, 86
%
\bibitem[Chen et al.(2022)]{Chen2022} Chen, Y.-C., Hwang, H.-C., Shen, Y., et al.\ 2022, \apj, 925, 162. doi:10.3847/1538-4357/ac401b
%
\bibitem[Chiba et al.(2005)]{Chiba2005} Chiba, M., Minezaki, T., Kashikawa, N., et al.\ 2005, \apj, 627, 53. doi:10.1086/430403
%
\bibitem[Comerford et al.(2009)]{Comerford2009} Comerford, J.~M., Griffith, R.~L., Gerke, B.~F., et al.\ 2009, \apjl, 702, L82
%
\bibitem[Comerford et al.(2012)]{Comerford2012} Comerford, J.~M., Gerke, B.~F., Stern, D., et al.\ 2012, \apj, 753, 42. doi:10.1088/0004-637X/753/1/42
%
\bibitem[Comerford et al.(2013)]{Comerford2013} Comerford, J.~M., Schluns, K., Greene, J.~E., et al.\ 2013, \apj, 777, 64
%
\bibitem[Comerford \& Greene(2014)]{Comerford2014} Comerford, J.~M. \& Greene, J.~E.\ 2014, \apj, 789, 112. doi:10.1088/0004-637X/789/2/112

\bibitem[Comerford et al.(2015)]{Comerford2015} Comerford, J.~M., Pooley, D., Barrows, R.~S., et al.\ 2015, \apj, 806, 219
%
\bibitem[De Angeli et al.(2022)]{DeAngeli2022}
De Angeli, F. , Weiler, M. , Montegriffo, P. , et al.\ 2022, arXiv/2206.06143 
%
\bibitem[Delchambre et al.(2019)]{Delchambre2019} Delchambre, L., Krone-Martins, A., Wertz, O., et al.\ 2019, \aap, 622, A165. doi:10.1051/0004-6361/201833802
%
\bibitem[Ding et al.(2017)]{Ding2017} Ding, X., Treu, T., Suyu, S.~H., et al.\ 2017, \mnras, 472, 90. doi:10.1093/mnras/stx1972
%
\bibitem[Ducourant et al.(2018)]{Ducourant2018} Ducourant, C., Wertz, O., Krone-Martins, A., et al.\ 2018, \aap, 618, A56. doi:10.1051/0004-6361/201833480
%
\bibitem[Eracleous et al.(2012)]{Eracleous2012} Eracleous, M., Boroson, T.~A., Halpern, J.~P., et al.\ 2012, \apjs, 201, 23. doi:10.1088/0067-0049/201/2/23
%
\bibitem[Feng et al.(2021)]{Feng2021} Feng, H.-C., Hu, C., Li, S.-S., et al.\ 2021, \apj, 909, 18. doi:10.3847/1538-4357/abd851
%
\bibitem[Fu et al.(2018)]{Fu2018} Fu, H., Steffen, J.~L., Gross, A.~C., et al.\ 2018, \apj, 856, 93. doi:10.3847/1538-4357/aab364
%
\bibitem[Gaia Collaboration et al.(2016)]{Gaia2016} \textit{Gaia} Collaboration, Prusti, T., de Bruijne, J.~H.~J., et al.\ 2016, \aap, 595, A1. doi:10.1051/0004-6361/201629272
%
\bibitem[Gaia Collaboration et al.(2018a)]{Gaia2018a} \textit{Gaia} Collaboration, Brown, A.~G.~A., Vallenari, A., et al.\ 2018, \aap, 616, A1. doi:10.1051/0004-6361/201833051
%
\bibitem[Gaia Collaboration et al.(2018b)]{Gaia2018b} \textit{Gaia} Collaboration, Mignard, F., Klioner, S.~A., et al.\ 2018, \aap, 616, A14. doi:10.1051/0004-6361/201832916
%
\bibitem[Gaia Collaboration et al.(2021)]{Gaia2021} \textit{Gaia} Collaboration, Brown, A.~G.~A., Vallenari, A., et al.\ 2021, \aap, 649, A1. doi:10.1051/0004-6361/202039657
%
\bibitem[Gaia Collaboration et al.(2022)]{Gaia2022} Gaia Collaboration, Bailer-Jones, C.~A.~L., Teyssier, D., et al.\ 2022, arXiv:2206.05681
%
\bibitem[Gaia Collaboration et al.(2022)]{GaiaCRF3} Gaia Collaboration, Klioner, S.~A., Lindegren, L., et al.\ 2022, \aap, 667, A148. doi:10.1051/0004-6361/202243483
%
\bibitem[Ge et al.(2012)]{Ge2012} Ge, J.-Q., Hu, C., Wang, J.-M., et al.\ 2012, \apjs, 201, 31
%
\bibitem[Graham(2004)]{Graham2004} Graham, A.~W.\ 2004, \apjl, 613, L33
%
\bibitem[Graham et al.(2015a)]{Graham2015a} Graham, M.~J., Djorgovski,S.~G., Stern, D., Glikman, E., Drake, A.~J., Mahaball, A.~A., Donalek,C., Larson, S., \& Christensen, E.\ 2015a, \nat, 518, 74
%
\bibitem[Graham et al.(2015b)]{Graham2015b} Graham, M.~J., Djorgovski, S.~G., Stern, D., et al.\ 2015b, \mnras, 453, 1562
%
\bibitem[Hwang et al.(2020)]{Hwang2020} Hwang, H.-C., Shen, Y., Zakamska, N., et al.\ 2020, \apj, 888, 73. doi:10.3847/1538-4357/ab5c1a
%
\bibitem[Inada et al.(2005)]{Inada2005} Inada, N., Burles, S., Gregg, M.~D., et al.\ 2005, \aj, 130, 1967. doi:10.1086/432930
%
\bibitem[Inada et al.(2007)]{Inada2007} Inada, N., Oguri, M., Becker, R.~H., et al.\ 2007, \aj, 133, 206. doi:10.1086/509702
%
\bibitem[Inada et al.(2008)]{Inada2008} Inada, N., Oguri, M., Becker, R.~H., et al.\ 2008, \aj, 135, 496. doi:10.1088/0004-6256/135/2/496
%
\bibitem[Inada et al.(2014)]{Inada2014} Inada, N., Oguri, M., Rusu, C.~E., et al.\ 2014, \aj, 147, 153. doi:10.1088/0004-6256/147/6/153
%
\bibitem[Jackson et al.(2015)]{Jackson2015} Jackson, N., Tagore, A.~S., Roberts, C., et al.\ 2015, \mnras, 454, 287. doi:10.1093/mnras/stv1982

\bibitem[Ji et al.(2021)]{Ji2021} Ji, X., Lu, Y., Ge, J., et al.\ 2021, \apj, 910, 101. doi:10.3847/1538-4357/abe386
%
\bibitem[J{\"o}nsson et al.(2020)]{Jonsson2020} J{\"o}nsson, H., Holtzman, J.~A., Allende Prieto, C., et al.\ 2020, \aj, 160, 120. doi:10.3847/1538-3881/aba592
%
\bibitem[Koss et al.(2012)]{Koss2012} Koss, M., Mushotzky, R., Treister, E., et al.\ 2012, \apjl, 746, L22
%
\bibitem[Krone-Martins et al.(2018)]{Krone-Martins2018} Krone-Martins, A., Delchambre, L., Wertz, O., et al.\ 2018, \aap, 616, L11. doi:10.1051/0004-6361/201833337
%
\bibitem[Krone-Martins et al.(2019)]{Krone-Martins2019} Krone-Martins, A., Graham, M.~J., Stern, D., et al.\ 2019, arXiv:1912.08977
%
\bibitem[Lacki et al.(2009)]{Lacki2009} Lacki, B.~C., Kochanek, C.~S., Stanek, K.~Z., et al.\ 2009, \apj, 698, 428. doi:10.1088/0004-637X/698/1/428
%
\bibitem[Leh{\'a}r et al.(2000)]{Lehar2006} Leh{\'a}r, J., Falco, E.~E., Kochanek, C.~S., et al.\ 2000, \apj, 536, 584. doi:10.1086/308963
%
\bibitem[Lemon et al.(2017)]{Lemon2017} Lemon, C.~A., Auger, M.~W., McMahon, R.~G., et al.\ 2017, \mnras, 472, 5023. doi:10.1093/mnras/stx2094
%
\bibitem[Lemon et al.(2018)]{Lemon2018} Lemon, C.~A., Auger, M.~W., McMahon, R.~G., et al.\ 2018, \mnras, 479, 5060. doi:10.1093/mnras/sty911
%
\bibitem[Lemon et al.(2019)]{Lemon2019} Lemon, C.~A., Auger, M.~W., \& McMahon, R.~G.\ 2019, \mnras, 483, 4242. doi:10.1093/mnras/sty3366
%
\bibitem[Lemon et al.(2022)]{Lemon2021} Lemon, C., Anguita, T., Auger, M., et al.\ 2022, arXiv:2206.07714

\bibitem[Li et al.(2016)]{Li2016} Li, Y.-R., Wang, J.-M., Ho, L.~C., et al.\ 2016, \apj, 822, 4
%
\bibitem[\protect\citeauthoryear{Liao et al.}{2019}]{Liao2019} Liao S.-L., Qi Z.-X., Guo S.-F., Cao Z.-H., 2019, RAA, 19, 029. doi:10.1088/1674-4527/19/2/29
%
\bibitem[Liao et al.(2021a)]{LiaoPASP2021a} Liao, S., Qi, Z., Cao, Z., et al.\ 2021a, \pasp, 133, 024501. doi:10.1088/1538-3873/abd4bd
%
\bibitem[Liao et al.(2021b)]{LiaoPASP2021b} Liao, S., Wu, Q., Qi, Z., et al.\ 2021b, \pasp, 133, 094501. doi:10.1088/1538-3873/ac1eeb
%
\bibitem[Liu et al.(2010a)]{Liu2010a} Liu, X., Shen, Y., Strauss, M.~A., et al.\ 2010, \apj, 708, 427. doi:10.1088/0004-637X/708/1/427
%
\bibitem[Liu et al.(2010b)]{Liu2010b} Liu, X., Greene, J.~E., Shen, Y., et al.\ 2010, \apjl, 715, L30
%
\bibitem[Liu et al.(2011)]{Liu2011} Liu, X., Shen, Y., Strauss, M.~A., et al.\ 2011, \apj, 737, 101
%
\bibitem[Liu et al.(2014)]{Liu2014} Liu, X., Shen, Y., Bian, F., et al.\ 2014, \apj, 789, 140
%
\bibitem[Liu et al.(2018)]{Liu2018} Liu, X., Guo, H., Shen, Y., et al.\ 2018, \apj, 862, 29
%
\bibitem[Lyke et al.(2020)]{Lyke2020} Lyke, B.~W., Higley, A.~N., McLane, J.~N., et al.\ 2020, \apjs, 250, 8. doi:10.3847/1538-4365/aba623
%
\bibitem[Mannucci et al.(2022)]{Mannucci2022} Mannucci, F., Pancino, E., Belfiore, F., et al.\ 2022, Nature Astronomy. doi:10.1038/s41550-022-01761-5
%
\bibitem[Magain et al.(1988)]{Magain1988} Magain, P., Surdej, J., Swings, J.-P., et al.\ 1988, \nat, 334, 325. doi:10.1038/334325a0
%
\bibitem[Makarov \& Secrest(2022)]{Makarov2022ApJ} Makarov, V.~V. \& Secrest, N.~J.\ 2022, \apj, 933, 28. doi:10.3847/1538-4357/ac7047
%
\bibitem[More et al.(2016)]{More2016} More, A., Oguri, M., Kayo, I., et al.\ 2016, \mnras, 456, 1595. doi:10.1093/mnras/stv2813
%
\bibitem[Morgan et al.(2001)]{Morgan2001} Morgan, N.~D., Becker, R.~H., Gregg, M.~D., et al.\ 2001, \aj, 121, 611. doi:10.1086/318744
%
\bibitem[Morokuma et al.(2007)]{Morokuma2007} Morokuma, T., Inada, N., Oguri, M., et al.\ 2007, \aj, 133, 214. doi:10.1086/509701
%
\bibitem[Oguri \& Keeton(2004)]{Oguri2004} Oguri, M. \& Keeton, C.~R.\ 2004, \apj, 610, 663. doi:10.1086/421870
%
\bibitem[Pindor et al.(2006)]{Pindor2006} Pindor, B., Eisenstein, D.~J., Gregg, M.~D., et al.\ 2006, \aj, 131, 41. doi:10.1086/497965
%
\bibitem[Popovi\'{c}(2012)]{Popovic2012} Popovi\'{c}, L. {\v{C}}.\ 2012, New Astro. Rev., 56, 74
%
\bibitem[Reimers et al.(2002)]{Reimer2002} Reimers, D., Hagen, H.-J., Baade, R., et al.\ 2002, \aap, 382, L26. doi:10.1051/0004-6361:20011798
%
\bibitem[Rodriguez et al.(2006)]{Rodriguez2006} Rodriguez, C., Taylor, G.~B., Zavala, R.~T., et al.\ 2006, \apj, 646, 49. doi:10.1086/504825
%
\bibitem[Schechter et al.(1998)]{Schechter1998} Schechter, P.~L., Gregg, M.~D., Becker, R.~H., et al.\ 1998, \aj, 115, 1371. doi:10.1086/300294
%
\bibitem[Sesana et al.(2009)]{Sesana2009} Sesana, A., Vecchio, A., \& Volonteri, M.\ 2009, \mnras, 394, 2255. doi:10.1111/j.1365-2966.2009.14499.x
%
\bibitem[Shen et al.(2011)]{Shen2011} Shen, Y., Liu, X., Greene, J.~E., et al.\ 2011, \apj, 735, 48
%
\bibitem[Shen et al.(2013)]{shen2013} Shen, Y., Liu, X., Loeb, A., et al.\ 2013, \apj, 775, 49
%
\bibitem[Shen et al.(2019)]{Shen2019} Shen, Y., Hwang, H.-C., Zakamska, N., et al.\ 2019, \apjl, 885, L4. doi:10.3847/2041-8213/ab4b54
%
\bibitem[Shen et al.(2021)]{Shen2021} Shen, Y., Chen, Y.-C., Hwang, H.-C., et al.\ 2021, Nature Astronomy, 5, 569. doi:10.1038/s41550-021-01323-1
%
\bibitem[Shen et al.(2022)]{shen2022} Shen, Y., Hwang, H.-C., Oguri, M., et al.\ 2022, arXiv:2208.04979
%
\bibitem[Silverman et al.(2020)]{Silverman2020} Silverman, J.~D., Tang, S., Lee, K.-G., et al.\ 2020, \apj, 899, 154. doi:10.3847/1538-4357/aba4a3
%
\bibitem[Sluse et al.(2011)]{Sluse2011} Sluse, D., Schmidt, R., Courbin, F., et al.\ 2011, \aap, 528, A100. doi:10.1051/0004-6361/201016110
%
\bibitem[Smith et al.(2010)]{Smith2010} Smith, K.~L., Shields, G.~A., Bonning, E.~W., et al.\ 2010, \apj, 716, 866. doi:10.1088/0004-637X/716/1/866
%
\bibitem[Souchay et al.(2022)]{Souchay2022} Souchay, J., Secrest, N., Lambert, S., et al.\ 2022, \aap, 660, A16. doi:10.1051/0004-6361/202141915
%
\bibitem[Stacey et al.(2018)]{Stacey2018} Stacey, H.~R., McKean, J.~P., Robertson, N.~C., et al.\ 2018, \mnras, 476, 5075. doi:10.1093/mnras/sty458
%
\bibitem[Suyu et al.(2017)]{Suyu2017} Suyu, S.~H., Bonvin, V., Courbin, F., et al.\ 2017, \mnras, 468, 2590. doi:10.1093/mnras/stx483
%
\bibitem[Tsalmantza et al.(2011)]{Tsalmantza2011} Tsalmantza, P., Decarli, R., Dotti, M., \& Hogg, D.~W.\ 2011, \apj, 738, 20
%
\bibitem[Volonteri et al.(2003)]{Volonteri2003} Volonteri, M., Haardt, F., \& Madau, P.\ 2003, \apj, 582, 559
%
\bibitem[Wu et al.(2021)]{Wu2021} Wu, Q., Liao, S., Qi, Z., et al.\ 2021, arXiv:2111.02131
%
\bibitem[Wu et al.(2022)]{Wu2022} Wu, Q.-Q., Liao, S.-L., Ji, X., et al.\ 2022, Frontiers in Astronomy and Space Sciences, 9, 822768. doi:10.3389/fspas.2022.822768
%
\bibitem[Yan et al.(2015)]{Yan2015} Yan, C.-S., Lu, Y., Dai, X., \& Yu, Q.\ 2015, \apj, 809, 117 %
%
\bibitem[Young et al.(1981)]{Young1981} Young, P., Deverill, R.~S., Gunn, J.~E., et al.\ 1981, \apj, 244, 723. doi:10.1086/158750

\bibitem[Yu et al.(2011)]{Yu2011} Yu, Q., Lu, Y., Mohayaee, R., et al.\ 2011, \apj, 738, 92
%
\bibitem[Yu(2002)]{Yu2002} Yu, Q.\ 2002, \mnras, 331, 935
%
\bibitem[Zheng et al.(2016)]{Zheng2016} Zheng, Z.-Y., Butler, N.~R., Shen, Y., et al.\ 2016, \apj, 827, 56 
%
\end{thebibliography}
\end{document}